\documentclass{article}

\usepackage{graphicx}
\usepackage{amsmath}
\usepackage{amssymb}
\usepackage{amsfonts}

\usepackage{enumitem}
\usepackage[latin1]{inputenc}

\usepackage{amsthm}
\usepackage{url}

\usepackage[left=3cm,right=3cm,top=3cm,bottom=3cm]{geometry}

\newtheoremstyle{ourstyle}
  {4mm}
  {3pt}
  {\upshape}
  {}
  {\bfseries}
  {.}
  {.5em}
  {}

\theoremstyle{definition}

\newtheorem{theorem}{Theorem}
\newtheorem{lemma}{Lemma}
\newtheorem{corollary}{Corollary}
\newtheorem{assumption}{Assumption}
\newtheorem{postulate}{Postulate}

\newtheorem*{comment}{Remark}

\newcounter{footnumber}
\newcommand{\dimostrazione}[3]{{\setcounter{footnumber}{\value{footnote}}\addtocounter{footnumber}{
1}\setcounter{footnote}{\value{footnumber}}The proof is  in Footnote \thefootnumber.\footnotetext{\textbf{Proof of #1 \ref{#2}.\ }#3\qed}\setcounter{footnumber}{\thefootnote}}}

\newcommand{\virgolettedue}[1]{{``#1''}}

\newcommand{\W}[3]{{{#1}_#2\xrightarrow{\rm w} {#1}_#3}}

\newcommand{\Wcomp}[2]{{{#1}_1#2_{\rm se1}\xrightarrow{\rm w} #1_2#2_{\rm se2}}}
\newcommand{\Wcomprev}[2]{{#1_1#2_{\rm se1}\xrightarrow{\rm wrev}#1_2#2_{\rm se2rev}}}

\newcommand{\FA}{{\textbf{F}^{A}_e}}

\newcommand{\RA}[1]{{\pmb{\textsl{\textsf{R}}}^A_{#1}}}
\newcommand{\RB}[1]{{\pmb{\textsl{\textsf{R}}}^B_{#1}}}
\newcommand{\RC}[1]{{\pmb{\textsl{\textsf{R}}}^C_{#1}}}
\newcommand{\RD}[1]{{\pmb{\textsl{\textsf{R}}}^D_{#1}}}
\newcommand{\pmbsub}[1]{{\mbox{\boldmath\scriptsize$#1$}}}

\newcommand{\spazio}{{\vskip 3pt\noindent}}

\title{A definition of thermodynamic entropy valid for\\ non-equilibrium states and few-particle systems}

\author{Gian Paolo Beretta$^{1}$ and Enzo Zanchini$^{2}$\\ \small
$^{1}$Universit\`{a} di Brescia, Dipartimento di
Ingegneria Meccanica e Industriale\\ \small Via Branze 38, 25123 Brescia,
Italy\\ \small
$^{2}$Universit\`{a} di Bologna, Dipartimento di
Ingegneria Energetica, Nucleare e del Controllo Ambientale\\ \small  Viale
Risorgimento 2, 40136 Bologna, Italy}

\begin{document}
\maketitle

\begin{abstract}
From a new rigorous formulation of the general axiomatic foundations of thermodynamics we derive an operational definition of entropy that responds to the emergent need in many technological frameworks to understand and deploy thermodynamic entropy well beyond the traditional realm of equilibrium states of macroscopic systems. The new definition is achieved by avoiding to resort to the traditional concepts of \virgolettedue{heat} (which restricts \emph{a priori} the traditional definitions of entropy to the equilibrium domain) and of \virgolettedue{thermal reservoir} (which restricts \emph{in practice} our previous definitions of non-equilibrium entropy to the many-particle domain). The measurement procedure that defines entropy is free from intrinsic limitations and can  be applied, \emph{in principle}, even to non-equilibrium states of few-particle systems, provided they are separable and uncorrelated. The construction starts from a previously developed set of carefully worded operational definitions for all the basic concepts. Then, through a new set of fully spelled-out fundamental hypotheses (four postulates and five assumptions) we derive the definitions of energy and entropy of any state, and of temperature of any stable equilibrium state. Finally, we prove the principle of entropy non-decrease, the  additivity of entropy differences, the maximum entropy principle, and the impossibility of existence of a thermal reservoir.
\end{abstract}


\section{Introduction}

Thermodynamic entropy plays a crucial role in the development of the physical foundations of a variety of emerging technologies --- nanomaterials, small-scale hydrodynamics, chemical kinetics for energy and environmental engineering and biotechnologies, electrochemistry, quantum entanglement in quantum information, non-equilibrium bulk and interface phenomena,  etc.\ ---  which require a clear understanding of the meaning and role of thermodynamic entropy beyond the traditional equilibrium and macroscopic realms, well into the non-equilibrium and few-particle domains currently being explored very actively in  many fields of science and technology (see, e.g., Refs.\ \cite{Horodecki13,Skrzypczyk14,Brandao13,Verley12} for recent attempts to extend thermodynamics to nonequilibrium states and individual quantum systems).

In traditional treatments of thermodynamics (see, e.g. Refs.\ \cite{Fermi37, Pippard57, Zemansky68}), the definitions of thermodynamic temperature and of entropy are based on the concepts of \emph{heat} and of \emph{thermal reservoir}. Usually, heat is not defined rigorously. For instance, in his lectures on physics, Feynman \cite{Feynman63} describes heat as one of several different forms of energy related to the jiggling motion of particles; in this picture, heat appears as a transfer of kinetic energy and the difference between heat and work is not clarified. Landau and Lifshitz \cite{Landau80} define heat as the part of an energy change of a body that is not due to work done on it. However, there are interactions between systems which are neither heat nor work, such as, for instance, exchanges of radiation between systems in nonequilibrium states. Guggenheim \cite{Guggenheim67} defines heat as an exchange of energy that differs from work and is determined by a temperature difference. Keenan \cite{Keenan41} defines heat as the energy transferred from one system to a second system at lower temperature, by virtue of the temperature difference, when the two are brought into communication. These definitions do not describe clearly the phenomena which occur at the interface between the interacting systems; moreover, they require a previous definition of \emph{empirical temperature}, a concept which, in turn, is usually not defined rigorously. Another drawback of the employment of heat in the definition of entropy is the following: since heat, when properly defined, requires the existence of subsystems in stable equilibrium at the boundary between the interacting systems, a definition of entropy based on heat can hold, at most, in the domain of \emph{local equilibrium} states.

An alternative method for the axiomatization of thermodynamics was developed at MIT by Hatsopoulos and Keenan \cite{HK} and by Gyftopoulos and Beretta \cite{GB}. The main progress obtained in these references, with respect to the traditional treatments, is a more general definition of entropy --- not based on the heuristic notions of empirical temperature and heat, and not restricted to stable equilibrium states --- that emerges from a complete set of operational definitions of the basic concepts, such as those of system, property, state, and stable equilibrium state, and a new statement of the second law expressed as a postulate of existence, for a system with fixed composition and constraints, of  a unique stable equilibrium state for each  value of the energy.

Improvements of this method, yielding more rigorous definitions of isolated system, environment of a system and external force field, as well as a more direct definition of entropy, have been proposed over the years by the present authors \cite{Zanchini86,Zanchini88,Zanchini92,ZBsymposium,ZBIJoT, BZInTech, ZBEntropy14}.  Such constructions are important because they provide rigorous operational definitions of non-equilibrium entropy. However, they still require the use of a thermal reservoir as an auxiliary system (that plays the role of an \textit{entropy meter}) in the operational procedure that defines how to measure the entropy difference between any two states of a system. As already pointed out in Ref.\ \cite[p.87]{GB}, such use of thermal reservoirs has both logical and operational drawbacks.

A thermal reservoir, when properly defined \cite{GB, ZBIJoT, BZInTech}, is a closed system $R$, contained in a fixed region of space, such that whenever $R$  is in stable equilibrium it is also in mutual stable equilibrium with a duplicate of itself, kept in any of its stable equilibrium states. Once thermodynamic temperature has been defined,  it turns out that a thermal reservoir has the same temperature in all its stable equilibrium states, independently of the value of the energy. This condition is fulfilled by the simple-system model\footnote{As defined and discussed in Ref.\ \cite[pp.263-265]{GB}, the simple-system model is appropriate for macroscopic systems with many particles, but fails for few-particle systems for which, e.g., rarefaction effects near walls cannot be neglected.} of a pure substance kept  in the range of triple-point stable equilibrium states, because within such range of states energy can be added or removed at constant volume without changing the temperature. Hence, pure substances in their triple-point ranges are good practical examples of thermal reservoirs that can be easily set up in any laboratory.

However, as  proved rigorously  in the present paper, for any closed system contained in a fixed region of space the temperature is a strictly increasing function of the energy. Therefore, the triple-point model is only an approximate description of reality, valid with exceedingly good approximation for systems with many particles of the order of one mole, but not in general, e.g., not for systems with few particles. In a fully explicit axiomatic treatment one could declare the existence of thermal reservoirs as an assumption, but then one could prove that, strictly, thermal reservoirs cannot exist.
Thus, from the strictly logical point of view, the use of the thermal reservoir in the definition of entropy  is an internal inconsistency. The scope of the present paper is to remove  such inconsistency, by developing new general definitions of thermodynamic temperature and thermodynamic entropy that are neither based on the concept of heat nor  on that of thermal reservoir.

Another important drawback of the use of a thermal reservoir $R$ in the measurement procedure that defines the entropy difference of two states $A_1$ and $A_2$ of a system $A$  is that the procedure \cite{GB, ZBIJoT, BZInTech,ZBEntropy14} requires to measure the energy change of the reservoir $R$  in a reversible weight process for the composite system $AR$ in which $A$ changes from state $A_1$ to state $A_2$. If system $A$ has only few particles, than the energy change of $R$ will be extremely small and hardly measurable if, as just discussed, the thermal reservoir $R$  can only be realized by means of a macroscopic system.

A procedure which yields the definitions of  temperature and  entropy without employing the concepts of heat and of thermal reservoir was presented by Constantin Carathéodory in 1909 \cite{Caratheodory09}. However, his treatment is valid only for simple systems in stable equilibrium states. The same restriction holds for some developments of Carathéodory's method \cite{Turner60,Landsberg61, Sears63,Giles64}, aimed at making the treatment simpler and less abstract.

Another axiomatization of thermodynamics has developed in recent years by Lieb and Yngvason \cite{LY, LY13, LY14}. Their method is based on establishing an order relation between states, denoted by the symbol $\prec $, through the concept of adiabatic accessibility: a state $Y$ is said to be adiabatically accessible from a state $X$, i.e., $X \prec Y$, if it is possible to change the state from $X$ to $Y$ by means of an adiabatic process. By introducing a suitable set of Axioms concerning the order relation $\prec$, the authors prove the existence and the \emph{essential uniqueness} \cite{LY} of entropy. While the treatment presented in Ref.\  \cite{LY} holds only for stable equilibrium states of simple systems or collections of simple systems, through the complements presented in Refs.\ \cite{LY13, LY14} the validity is extended respectively to non-equilibrium states  \cite{LY13} and, through the use of a simple system as an \emph{entropy meter}, also to non-simple systems \cite{LY14}. Since to exhibit simple-system behavior the entropy meter must be a many-particle system, when applied  to few-particle systems the definition could present the same kind of 'practical'  problems faced by our previous definitions based on the entropy meter being a thermal reservoir.

In the present paper, a set of postulates and assumptions analogous to that stated in Ref. \cite{ZBEntropy14} is employed, but here the definitions of thermodynamic temperature and thermodynamic entropy are obtained \textit{without employing the concept of thermal reservoir}. Indeed,  to point out that the use of thermal reservoirs is, strictly speaking, a logical inconsistency, we prove by a theorem the impossibility of existence of a thermal reservoir. The main result of the new formulation is that by avoiding to use as entropy meter a many-particle system, we derive a rigorous and general operational definition of thermodynamic entropy which holds, potentially, also for non-equilibrium states of non-simple and non-macroscopic systems.

The potential applicability to non-equilibrium states is a relevant feature in the framework of the fast growing field of non-equilibrium thermodynamics (see, e.g., Ref. \cite{Kjelstrup08}), where research advances seem to substantiate from many perspectives the validity of a general principle of maximum entropy production \cite{Gheorghiu01,Gheorghiu01add,Martyushev06,Beretta09,Beretta14}.

The potential applicability to non-macroscopic systems is a relevant feature in the framework of the recently growing field of  thermodynamics in the quantum regime, where much discussion about the microscopic foundations of thermodynamics  is still taking place (see, e.g., Ref.\ \cite{Maddox85,
Beretta86,
Hatsopoulos08,Bennett08,Lloyd08,Maccone11,Horodecki13,Skrzypczyk14,Brandao13,Verley12}).

The definition of entropy presented here is complementary to that developed by Lieb and Yngvason: indeed, while Refs.\ \cite{LY, LY13, LY14} are focused on the proof of existence and essential uniqueness of an entropy function which is additive and fulfils the principle of entropy nondecrease, the present treatment identifies a general measurement procedure suitable to determine the entropy values.

In order to focus immediately on the construction of the new general definition of entropy, we keep to a minimum the discussion of the preliminary concepts.  Therefore, only a brief summary of the basic definitions is presented here, because a complete set of operational definitions is available in Refs.\ \cite{ZBIJoT,BZInTech}. Instead, we provide in footnotes full proofs of the lemmas, theorems, and corollaries.

\section{Summary of basic definitions}
With the term \textit{system} we mean a set of material particles, of one or more kinds, such that, at each instant of time, the particles of each kind are contained within a given region of space. If the boundary surfaces of the regions of space which contain the particles of the systems are all \textit{walls},\textit{ i.e.}, surfaces which cannot be crossed by material particles, the system is called \textit{closed}. Any system is endowed with a set of reproducible measurement procedures; each procedure defines a \textit{property} of the system. The set of all the values of the properties of a system, at a given instant of time, defines the \textit{state} of the system at that instant.\\
A system can be in contact with other matter, or surrounded by empty space; moreover, force fields due to external matter can act in the region of space occupied by the system. If, at an instant of time, all the particles of the system are removed from the respective regions of space and brought far away, but a force field is still present in the region of space (previously) occupied by the system, then this force field is called an \textit{external force field}. An external force field can be either gravitational, or electric or magnetic, or a superposition of the three.\\
Consider the union of all the regions of space spanned by a system during its entire time evolution. If no other material particles, except those of the system, are present in the region of space spanned by the system or touches the boundary of this region, and if the external force field in this region is either vanishing or stationary, then we say that the system is \textit{isolated}. Suppose that an isolated system $I$ can be divided into two subsystems, $A$ and $B$. Then, we can say that $B$ is the \textit{environment} of $A$ and viceversa.\\
If, at a given instant of time, two systems $A$ and $B$ are such that the force field produced by $B$ is vanishing in the region of space occupied by $A$ and viceversa, then we say that $A$ and $B$ are \textit{separable} at that instant. The energy of a system $A$ is defined (see Section 3) only for the states of $A$ such that $A$ is separable from its environment. Consider, for instance, the following simple example from mechanics. Let $A$ and $B$ be rigid bodies in deep space, far away from any other object and subjected to a mutual gravitational force. Then, the potential energy of the composite system $AB$ is defined, but that of $A$ and of $B$ is not.\\
If, at a given instant of time, two systems $A$ and $B$ are such that the outcomes of the measurements performed on $B$ are statistically independent of those of the measurements performed on $A$, and viceversa, we say that $A$ and $B$ are \textit{uncorrelated from each other} at that instant. The entropy of a system $A$ is defined (see Section 5) only for the states of $A$ such that $A$ is separable and uncorrelated from its environment.\\
We call \textit{process} of a system $A$ from state $A_1$ to state $A_2$ the time evolution of the isolated system $AB$ from $(AB)_1$ (with $A$ in state $A_1$) to $(AB)_2$ (with $A$ in state $A_2$), where $B$ is the environment of $A$. A process of $A$ is \textit{reversible} if the isolated system $AB$ can undergo a time evolution which restores it in its initial state $(AB)_1$. A process of a system $A$ is called a \textit{cycle} for $A$ if the final state $A_2$ coincides with the initial state $A_1$. A cycle for $A$ is not necessarily a cycle for $AB$.\\
An \textit{elementary mechanical system} is a system such that the only admissible change of state for it is a space translation in a uniform external force field; an example is a particle which can only change its height in a uniform external gravitational field. A process of a system $A$ from state $A_1$ to $A_2$, such that both in $A_1$ and in $A_2$ system $A$ is separable from its environment, is a \textit{weight process} for $A$ if the only net effect of the process in the environment of $A$ is the change of state of an elementary mechanical system.\\
An \textit{equilibrium} state of a system is a state such that the system is separable, the state does not vary with time, and it can be reproduced while the system is isolated. An
equilibrium state of a closed system $A$ in which $A$ is uncorrelated from its environment $B$, is called a \textit{stable equilibrium state} if it cannot be modified by any process between states in which $A$ is separable and uncorrelated from its environment such that neither  the geometrical configuration of the walls which bound the regions of space $\RA{}$ where the constituents of $A$ are contained, nor the state of the environment $B$ of $A$ have net changes. Two systems, $A$ and $B$, are in \textit{mutual stable equilibrium} if the composite system $AB$ (\textit{i.e.}, the union of both systems) is in a stable equilibrium state.
%

\section{Definition of energy for a closed system}

\subsection*{Weight polygonal and work in a weight polygonal.} Consider an ordered set of \emph{n} states of a closed system $A$, $(A_1, A_2, ... , A_n)$, such that in each of these states $A$ is separable from its environment. If \emph{n} - 1 weight processes exist, which interconnect $A_1$ and $A_2$, ... , $A_{n-1}$ and $A_n$, regardless of the direction of each process, we say that $A_1$ and $A_n$ can be interconnected by a weight polygonal.
For instance, if weight processes $A_1\xrightarrow{\rm w}A_2$ and $A_3\xrightarrow{\rm w}A_2$ exist for $A$,
we say that $A_1\xrightarrow{\rm w}A_2\xleftarrow{\rm w}A_3$ is a weight polygonal for $A$ from $A_1$ to $A_3$. We call work done by $A$ in a weight polygonal from $A_1$ to $A_n$ the sum of the works done by $A$ in the weight processes with direction from $A_1$ to $A_n$ and the opposites of the works done by $A$ in the weight processes with direction from $A_n$ to $A_1$ [15]. The work done by $A$ in a weight polygonal from $A_1$ to $A_n$ will be denoted by $W_{1n}^{A\xrightarrow{\rm wp}}$ ; its opposite will be called work received by $A$ in a weight polygonal from $A_1$ to $A_n$ and will be denoted by $W_{1n}^{A\xleftarrow{\rm wp}}$. Clearly, for a given weight polygonal, $W_{1n}^{A\xleftarrow{\rm wp}} = - W_{1n}^{A\xrightarrow{\rm wp}} = W_{n1}^{A\xrightarrow{\rm wp}}$. For the example of weight polygonal $A_1\xrightarrow{\rm w}A_2\xleftarrow{\rm w}A_3$  considered above, we have
\begin{equation}\label{workinpolygonal}
W_{13}^{A\xrightarrow{\rm wp}} = W_{12}^{A\rightarrow} - W_{32}^{A\rightarrow} .
\end{equation}

\begin{assumption}\label{Assumption1} Every pair of states ($A_1$, $A_2$) of a closed system $A$, such that $A$ is separable from its environment in both states, can be interconnected by means of a weight polygonal for $A$.
\end{assumption}

\begin{postulate}\label{Postulate1} The works done by a system in any two weight polygonals between the same initial and final states are identical.
\end{postulate}

\begin{comment} In Ref. \cite{Zanchini86} it is proved that, in sets of states where sufficient conditions of interconnectability by weight processes hold, Postulate \ref{Postulate1} can be proved as a consequence of the traditional form of the First Law, which concerns weight processes (or adiabatic processes).
\end{comment}

\subsection*{Definition of energy for a closed system. Proof
that it is a property.} Let ($A_1$, $A_2$) be any pair of states
of a system $A$, such that $A$ is separable from its environment in both states. We call \emph{energy difference} between
states $A_2$ and $A_1$ the work received by $A$ in any weight polygonal from $A_1$ to $A_2$, expressed as
\begin{equation}\label{energy}
E^A_2 - E^A_1 = - W_{12}^{A\xrightarrow{\rm wp}} = W_{12}^{A\xleftarrow{\rm wp}}.
\end{equation}
The First Law yields the following consequences: \\
(\emph{a}) the energy difference between two states $A_2$ and $A_1$ depends
only on the states $A_1$ and $A_2$;\\
(\emph{b}) (\emph{additivity of energy differences}) consider a pair of states $(AB)_1$
and $(AB)_2$ of a composite system $AB$, and denote by $A_1, B_1$ and $A_2, B_2$ the corresponding states of $A$ and $B$; then, if $A$, $B$ and $AB$ are separable from their environment in the states considered,
\begin{equation}\label{additivity}
E^{AB}_2 - E^{AB}_1 = E^A_2 - E^A_1 + E^B_2 - E^B_1 \;\; ;
\end{equation}
(\emph{c}) (\emph{energy is a property}) let $A_0$ be a reference
state of a system $A$, in which $A$ is separable from its environment, to which we assign an arbitrarily chosen value of energy $E^A_0$; the value of the energy of $A$ in any other state $A_1$ in which $A$ is separable from its environment is determined uniquely by
\begin{equation}\label{energyabs}
E^A_1 = E^A_0 + W_{01}^{A\xleftarrow{\rm wp}} \;\; ,
\end{equation}
where $W_{01}^{A\xleftarrow{\rm wp}}$ is the work received by $A$ in any weight polygonal for $A$ from $A_0$ to $A_1$.\\ Simple proofs of these consequences can be found in Ref. \cite{Zanchini86}, and will not be repeated here.

\section{Definition of temperature of a stable equilibrium state}

\begin{postulate}\label{Postulate2} Among all the states of a system $A$ such that the constituents of $A$ are contained in a given set of regions of space $\RA{}$, there is a stable equilibrium state for every value of the energy
$E^A$.
\end{postulate}

\begin{assumption}\label{Assumption2} Starting from any state in which the system is separable from its environment, a closed system $A$ can be changed to a stable equilibrium state with the same energy by means of a zero work weight process for $A$ in which the regions of space occupied by the constituents of $A$ have no net changes.
\end{assumption}

\begin{lemma}\label{Lemma1}
\textbf{Uniqueness of the stable equilibrium
state  for a given value of the energy}. There can be no pair of
different stable equilibrium states of a closed system $A$ with
identical regions of space $\RA{}$ and the same value of the
energy $E^A$. \dimostrazione{Lemma}{Lemma1}{Since $A$ is closed and in any stable equilibrium state it is separable and uncorrelated from its environment, if two such states existed, by Assumption \ref{Assumption2} the system could be changed from one to the other by means of a zero-work weight process, with no change of the regions of space occupied by the constituents of $A$ and no change of the state of the environment of $A$. Therefore, neither would satisfy the definition of stable equilibrium state.}
\end{lemma}

\begin{postulate}\label{Postulate3} There exist systems, called normal systems, whose energy has no upper bound. Starting from any state in which the system is separable from its environment, a normal system $A$ can be changed to a non-equilibrium state with arbitrarily higher energy (in which $A$ is separable from its environment) by means of a weight process for $A$ in which the regions of space occupied by the constituents of $A$ have no net changes. \end{postulate}

\begin{comment} The additivity of energy implies that the union of two or more normal systems, each separable from its environment, is a normal system to which Postulate \ref{Postulate3} applies.\\
In traditional treatments of thermodynamics only normal systems are considered, without an
explicit mention of this restriction. Moreover, Postulate \ref{Postulate3} is
\textit{not stated, but it is used}, for example in theorems where
one says that any amount of work can be transferred to a thermal
reservoir by a stirrer. Any system whose constituents have
translational, rotational or vibrational degrees of freedom is a
normal system.\\ On the other hand, quantum theoretical model
systems, such as spins, qubits, qudits, etc., whose energy is
bounded also from above, are special systems.
\end{comment}

\subsection*{Restriction to Normal Closed Systems.} In this paper, to focus the
attention of the reader on the main result of the
paper, namely avoiding the use of the concept of thermal reservoir
in the foundations of thermodynamics, we consider only normal
closed systems. The extension of the treatment to \textit{special
systems} and \textit{open systems} will
be presented elsewhere.

\begin{theorem}\label{Theorem1}
\textbf{Impossibility of a Perpetual Motion Machine of the Second Kind (PMM2)}. If  a normal system $A$ is in a stable equilibrium state, it is impossible to lower its energy by means of a weight process for $A$ in which the regions of space occupied by the constituents of $A$ have no net change. \dimostrazione{Theorem}{Theorem1}{
Suppose that, starting from a stable equilibrium state $A_{se}$ of
$A$, by means of a weight process $\Pi_1$ with positive work
$W^{A\rightarrow}=W>0$, the energy of $A$ is lowered and the
regions of space $\RA{}$ occupied by the constituents of $A$ have no net change. On account of Postulate \ref{Postulate3}, it would be possible to perform a weight process $\Pi_2$ for $A$ in which its regions of space $\RA{}$  have no net
change, the weight $M$ is restored to its initial state so that
the positive amount of energy $W^{A\leftarrow}=W>0$ is supplied
back to $A$, and the final state of $A$ is a non-equilibrium state,
namely, a state clearly different from $A_{se}$. Thus, the
composite zero-work weight process ($\Pi_1$, $\Pi_2$) would
violate the definition of stable equilibrium state. }
\end{theorem}



\begin{comment} \textit{Kelvin-Planck statement of the Second Law}.
As noted in Refs.\ \cite{HK} and \cite[p.64]{GB}, the impossibility
of a PMM2, which is
also known as the \textit{Kelvin-Planck statement of the Second
Law}, is a corollary of the definition of stable equilibrium
state, provided that we adopt the (usually implicit) restriction
to normal systems.
\end{comment}

\subsection*{Weight process for $AB$, standard with respect to
$B$.} Given a pair of states $(A_1, A_2)$ of a system $A$, such that $A$ is separable from its environment, and a system $B$ in the environment of $A$, we call \textit{weight process for $AB$, standard with
respect to $B$}  a weight process
$A_1B_{\textrm{se}1}\xrightarrow{\rm w} A_2B_{\textrm{se}2}$ for
the composite system $AB$ in which the end states of $A$ are the
given states $A_1$ and $A_2$, and the end states of $B$ are stable
equilibrium states with identical regions of space $\RB{}$.   For
a weight process for $AB$, standard with respect to $B$, we denote
the final energy of system $B$ by the symbol $E^B_{\rm se2}\big|_{A_1 A_2}^{{\rm sw,}B_{\rm se1}}$
or, if the process is reversible, $E^B_{\rm se2rev}\big|_{A_1 A_2}^{{\rm sw,}B_{\rm se1}}$ (when the context allows it, we simply denote them by $E^B_{\rm se2}$ and $E^B_{\rm se2rev}$, respectively).

\begin{comment} The term \virgolettedue{standard with respect to
$B$} is a shorthand to express the conditions that: 1) the end
states of $B$ are stable equilibrium, and 2) the regions of space
$\RB{\rm se1}$ and $\RB{\rm se2}$ are identical. The regions of
space $\RA{1}$ and $\RA{2}$, instead, need not be identical.
\end{comment}

\begin{assumption}\label{Assumption3} For
any given pair of states ($A_1$, $A_2$) of any closed system $A$ such that $A$ is separable and uncorrelated from its environment, it is always possible to find or to include in the environment of $A$ a system $B$ which has a stable equilibrium state $B_{\textrm{se}1}$ such
that the states $A_1$ and $A_2$ can be interconnected by means of
a reversible weight process for  $AB$, standard with respect to
$B$, in which system $B$ starts from state $B_{\textrm{se}1}$.
\end{assumption}

\begin{comment} Since Assumptions \ref{Assumption1} and \ref{Assumption2} can be considered as having a completely general validity, one can state that the domain of validity of Assumption \ref{Assumption3}
coincides with that of the definition of entropy given in this paper. If Assumption \ref{Assumption3} held for every pair of states of every closed system $A$ in which $A$ is separable and uncorrelated from its environment, including local non-equilibrium states of few-particle systems, then the definition of entropy given in this paper would be completely general.\\
If, for a given pair of states ($A_1$, $A_2$), a stable equilibrium state $B_{\textrm{se}1}$ of $B$ fulfills Assumption \ref{Assumption3}, then any other stable equilibrium state of $B$ with the same regions of space and
with an energy value higher than that of $B_{\textrm{se}1}$
fulfills Assumption \ref{Assumption3}, as well. Therefore, for a given pair of states ($A_1$, $A_2$) of a system $A$ and a selected system
$B$, there exists infinite different choices for
$B_{\textrm{se}1}$.\end{comment}

\begin{theorem}\label{Theorem2}
Given a pair of states ($A_1$, $A_2$)
of a system $A$ such that $A$ is separable and uncorrelated from its environment, a system $B$ in the environment of $A$, and an initial stable
equilibrium state $B_{\textrm{se}1}$, among all the weight
processes for $AB$, standard with respect to $B$, in which $A$
goes from $A_1$ to $A_2$ and $B$ begins in state
$B_{\textrm{se}1}$, the energy $E^B_{\rm se2}\big|_{A_1 A_2}^{{\rm sw,}B_{\rm se1}}$ of system $B$ in its final state
has a lower bound, $E^B_{\rm se2rev}\big|_{A_1 A_2}^{{\rm sw,}B_{\rm se1}}$, which is reached if and only if the process is
reversible. Moreover, for all such reversible processes, system
$B$ ends in the same stable equilibrium state
$B_{\textrm{se}2\textrm{rev}}$. \dimostrazione{Theorem}{Theorem2}{Consider a weight process for $AB$,
standard with respect to $B$, $\Pi_{AB}=\Wcomp{A}{B}$, a
reversible weight processes for $AB$, standard with respect to
$B$, $\Pi_{AB\rm rev}=\Wcomprev{A}{B}$, and the corresponding final
energies  of $B$, respectively, $E^B_{\rm se2}\big|_{A_1 A_2}^{{\rm sw,}B_{\rm se1}}$ and $E^B_{\rm se2rev}\big|_{A_1 A_2}^{{\rm sw,}B_{\rm se1}}$. We will prove that:
\\(a) $E^B_{\rm se2rev}\big|_{A_1 A_2}^{{\rm sw,}B_{\rm se1}} \leq E^B_{\rm se2}\big|_{A_1 A_2}^{{\rm sw,}B_{\rm se1}}$;
\\ (b) if also $\Pi_{AB}$ is reversible, then
$E^B_{\rm se2}\big|_{A_1 A_2}^{{\rm sw,}B_{\rm se1}} = E^B_{\rm se2rev}\big|_{A_1 A_2}^{{\rm sw,}B_{\rm se1}}$, and the end stable equilibrium state of $B$ is
the same, \textit{ i.e.}, $B_{\textrm{se}2}=B_{\textrm{se}2\textrm{rev}}$;\\
(c) if $E^B_{\rm se2}\big|_{A_1 A_2}^{{\rm sw,}B_{\rm se1}} = E^B_{\rm se2rev}\big|_{A_1 A_2}^{{\rm sw,}B_{\rm se1}}$, then also $\Pi_{AB}$ is
reversible.
\spazio\textit{Proof of (a)}.  Let us suppose, \textit{ab absurdo},
that  the energy of $B$ in
state $B_{\textrm{se}2}$ is lower than that in state
$B_{\textrm{se}2\textrm{rev}}$. Then, the composite process
($-\Pi_{AB\rm rev}$, $\Pi_{AB}$)
would be a weight process for $B$ in which, starting from the stable equilibrium state $B_{\textrm{se}2\textrm{rev}}$, the energy of $B$ is lowered and
its regions of space  have no net changes, in contrast with
Theorem \ref{Theorem1}. Therefore, $E^B_{\rm se2rev}\big|_{A_1 A_2}^{{\rm sw,}B_{\rm se1}} \leq E^B_{\rm se2}\big|_{A_1 A_2}^{{\rm sw,}B_{\rm se1}}$.

\spazio \textit{Proof of (b)}. If also process $\Pi_{AB}$ is
reversible, then, in addition to $E^B_{\rm se2rev}\big|_{A_1 A_2}^{{\rm sw,}B_{\rm se1}} \leq E^B_{\rm se2}\big|_{A_1 A_2}^{{\rm sw,}B_{\rm se1}}$,
also the relation $E^B_{\rm se2}\big|_{A_1 A_2}^{{\rm sw,}B_{\rm se1}} \leq E^B_{\rm se2rev}\big|_{A_1 A_2}^{{\rm sw,}B_{\rm se1}}$ must hold by
virtue of the proof of a) just given and, therefore, $E^B_{\rm se2rev}\big|_{A_1 A_2}^{{\rm sw,}B_{\rm se1}} = E^B_{\rm se2}\big|_{A_1 A_2}^{{\rm sw,}B_{\rm se1}}$. On account of Postulate \ref{Postulate2} and Lemma \ref{Lemma1}, the final
value of the energy of $B$ determines a unique final stable
equilibrium state of $B$; therefore
$B_{\textrm{se}2}=B_{\textrm{se}2\textrm{rev}}$.

\spazio \textit{Proof of (c)}. Let $\Pi_{AB}$ be  such that
$E^B_{\rm se2}\big|_{A_1 A_2}^{{\rm sw,}B_{\rm se1}} = E^B_{\rm se2rev}\big|_{A_1 A_2}^{{\rm sw,}B_{\rm se1}}$.  Then, the final states
$B_{\textrm{se}2}$ and $B_{\textrm{se}2\textrm{rev}}$ have the
same energy and, being stable equilibrium states, by Lemma \ref{Lemma1} they
must coincide. Thus, the composite process ($\Pi_{AB}$,
$-\Pi_{AB\rm rev}$) is a cycle for the isolated system $ABC$,
where $C$ is the environment of $AB$, where the only effect is the
return of the weight to its initial position. As a consequence,
being a part of a cycle of the isolated system $ABC$, process
$\Pi_{AB}$ is reversible.}
\end{theorem}

\begin{theorem}\label{Theorem3}
Consider a pair of states ($A_1$, $A_2$)
of a system $A$ such that $A$ is separable and uncorrelated from its environment, and two systems in the environment of $A$, $B$ and $C$, in given initial stable
equilibrium states $B_{\textrm{se}1}$ and $C_{\textrm{se}1}$. Let $\Pi_{AB\textrm{rev}}$ and $\Pi_{AC\textrm{rev}}$ be reversible weight processes
for $AB$ and for $AC$, both from $A_1$ to $A_2$ and standard with respect to $B$ and $C$ respectively, with the given initial states $B_{\textrm{se}1}$ and $C_{\textrm{se}1}$ of $B$ and $C$; let $E^B$ denote, for shorthand, the final energy  $E^B_{\rm se2rev}\big|_{A_1 A_2}^{{\rm sw,}B_{\rm se1}}$ of $B$ in process $\Pi_{AB\textrm{rev}}$ and let $E^C$ denote the final energy  $E^C_{\rm se2rev}\big|_{A_1 A_2}^{{\rm sw,}C_{\rm se1}}$ of $C$ in process $\Pi_{AC\textrm{rev}}$. Then, if $E^B-E^B_{\textrm{se}1}$ is vanishing, $E^C-E^C_{\textrm{se}1}$ is vanishing as well; if $E^B-E^B_{\textrm{se}1}$ is non vanishing, $E^C-E^C_{\textrm{se}1}$ is non vanishing and the ratio $(E^B-E^B_{\textrm{se}1}) \big / (E^C-E^C_{\textrm{se}1})$ is positive. \dimostrazione{Theorem}{Theorem3}{Assume that $E^B-E^B_{\textrm{se}1}$ is vanishing and that $E^C-E^C_{\textrm{se}1}$ is positive, and consider the composite process
($\Pi_{AB\textrm{rev}}, -\Pi_{AC\textrm{rev}}$);
this would be a reversible weight process for $C$ in which the energy change of $C$ is negative, the regions of space occupied by $C$ did not change and the initial state of $C$ is a stable equilibrium state, in contrast with Theorem \ref{Theorem1}. Assume now that $E^B-E^B_{\textrm{se}1}$ is vanishing and that $E^C-E^C_{\textrm{se}1}$ is negative, and consider the composite process ($\Pi_{AC\textrm{rev}},  -\Pi_{AB\textrm{rev}}$); this would be a reversible weight process for $C$ in which the energy change of $C$ is negative, the regions of space occupied by $C$ do not change and the initial state of $C$ is a stable equilibrium state, in contrast with Theorem \ref{Theorem1}. Then, if $E^B-E^B_{\textrm{se}1}$ is vanishing, $E^C-E^C_{\textrm{se}1}$ is vanishing as well.\\
Assume now that the energy change of $B$
is negative, \textit{i.e.}, $E^B-E^B_{\textrm{se}1}
<0$. Clearly, the energy change of $C$ cannot be
zero, because this would imply $E^B-E^B_{\textrm{se}1}=0$. Suppose that the energy change of $C$ is positive, $E^C-E^C_{\textrm{se}1}>0$, and consider the composite process ($\Pi_{AB\textrm{rev}}, -\Pi_{AC\textrm{rev}}$). In this process, which is a cycle for $A$, system $BC$ would have performed a positive work, given (energy balance for $BC$) by the sum of two positive addenda, namely $W = -(E^B-E^B_{\textrm{se}1}) +(E^C-E^C_{\textrm{se}1})$. On account of Postulate \ref{Postulate3} and Assumption \ref{Assumption2}, one could supply back to system $C$ a positive work amount equal to $(E^C-E^C_{\textrm{se}1})$ and restore $C$ to its initial state $C_{\rm se1}$
by means of a composite weight process $\Pi_C=C_{\rm
se2}\xrightarrow{\rm w}C_{\rm 3}\xrightarrow{\rm w}C_{\rm se1}$
where $C_3$ has energy $E^C_3=E^C_{\rm se1}$. Thus, the composite
process $(\Pi_{AB\textrm{rev}}, -\Pi_{AC\textrm{rev}}, \Pi_C)$ would be a again a weight
process for $B$ which violates Theorem \ref{Theorem1}. Therefore, if $E^B-E^B_{\textrm{se}1}$ is negative, $E^C-E^C_{\textrm{se}1}$ is negative as well.\\
Let us assume now that, in process $\Pi_{AB\rm rev}$, the energy change of $B$ is
positive. Then, in the reverse process $- \Pi_{AB\rm rev}$, the
energy change of $B$ is negative and, as we have just proved, the
energy change of $C$ in the reverse process - $\Pi_{AC\rm rev}$ must be negative as well. Therefore, in process $\Pi_{AC\rm rev}$, the energy change of $C$ is positive.}
\end{theorem}

\begin{lemma}\label{Lemma3}
Consider a pair of systems, $B$ and $C$, a pair of stable equilibrium states of these systems, $B_{\textrm{se}1}$ and $C_{\textrm{se}1}$, and a system $X$ in the environment of $BC$ with an initial state $X_1$ such that: every stable equilibrium state of $B$ with the same regions of space as $B_{\textrm{se}1}$ can be interconnected with $B_{\textrm{se}1}$ by a reversible weight process for $XB$ starting from $(X_1,B_{\textrm{se}1})$; every stable equilibrium state of $C$ with the same regions of space as $C_{\textrm{se}1}$ can be interconnected with $C_{\textrm{se}1}$ by a reversible weight process for $XC$ starting from $(X_1,C_{\textrm{se}1})$.\\ Denote by $\{(\Pi^{B_{\textrm{se}1}}_{XB\textrm{rev}}; \Pi^{C_{\textrm{se}1}}_{XC\textrm{rev}})\}$ the set of all the pairs of reversible weight processes for $XB$ and for $XC$, standard with respect to $B$ and $C$ and with initial states $(X_1, B_{\textrm{se}1})$ and $(X_1, C_{\textrm{se}1})$ respectively. Let $\{(B_{\textrm{se}2};C_{\textrm{se}2})\}$ the set of pairs  of final states of $B$ and $C$ which one obtains by the set of pairs of processes $\{(\Pi^{B_{\textrm{se}1}}_{XB\textrm{rev}}; \Pi^{C_{\textrm{se}1}}_{XC\textrm{rev}})\}$, and let $\{(E^B_{\textrm{se}2};E^C_{\textrm{se}2})\}$ be the corresponding values of the energy of $B$ and $C$. Then, the set of pairs of processes $\{(\Pi^{B_{\textrm{se}1}}_{XB\textrm{rev}}; \Pi^{C_{\textrm{se}1}}_{XC\textrm{rev}})\}$ determines a single valued and invertible function from the set $\{E^B_{\textrm{se}2}\}$ to the set $\{E^C_{\textrm{se}2}\}$,
\begin{equation}\label{function}
E^C = f^{B\rightarrow C}_{11}( E^B)\;\; ,
\end{equation}
which is independent on the choice of system $X$ and on the initial state $X_1$ used to construct the set of processes $\{(\Pi^{B_{\textrm{se}1}}_{XB\textrm{rev}}; \Pi^{C_{\textrm{se}1}}_{XC\textrm{rev}})\}$. \dimostrazione{Lemma}{Lemma3}{Choose a system $X$ and an initial state $X_1$ of $X$, and consider a pair of reversible weight processes $(\Pi^{B_{\textrm{se}1}}_{XB\textrm{rev}}; \Pi^{C_{\textrm{se}1}}_{XC\textrm{rev}})$, which belongs to the set $\{(\Pi^{B_{\textrm{se}1}}_{XB\textrm{rev}}; \Pi^{C_{\textrm{se}1}}_{XC\textrm{rev}})\}$. Let $X_2$, $B_{\textrm{se}2}$ and $C_{\textrm{se}2}$ be the final states of $X$, $B$ and $C$ for this pair of processes. Choose now a system $X'$ and an initial state $X'_1$ of $X'$, and consider a pair of reversible weight processes $(\Pi^{B_{\textrm{se}1}}_{X'B\textrm{rev}}; \Pi^{C_{\textrm{se}1}}_{X'C\textrm{rev}})$, which belongs to the set $\{(\Pi^{B_{\textrm{se}1}}_{X'B\textrm{rev}}; \Pi^{C_{\textrm{se}1}}_{X'C\textrm{rev}})\}$. Let $X'_2$, $B_{\textrm{se}3}$ and $C_{\textrm{se}3}$ be the final states of $X$, $B$ and $C$ for this pair of processes. We will prove that, if $B_{\textrm{se}3}$ coincides with $B_{\textrm{se}2}$, then also $C_{\textrm{se}3}$ coincides with $C_{\textrm{se}2}$, so that the correspondence between the final stable equilibrium states of $B$ and $C$ is not affected by either the choice of the auxiliary system, $X$ or $X'$, or the choice of the initial state of the auxiliary system.\\
Consider the composite system $XX'BC$, in the initial state $X_1X'_2B_{\textrm{se}1}C_{\textrm{se}2}$, and consider the composite process $\Pi=(\Pi^{B_{\textrm{se}1}}_{XB\textrm{rev}}, - \Pi^{C_{\textrm{se}1}}_{XC\textrm{rev}}, - \Pi^{B_{\textrm{se}1}}_{X'B\textrm{rev}},  \Pi^{C_{\textrm{se}1}}_{X'C\textrm{rev}})$, where $- \Pi^{C_{\textrm{se}1}}_{XC\textrm{rev}}$ is the reverse of $\Pi^{C_{\textrm{se}1}}_{XC\textrm{rev}}$ and $- \Pi^{B_{\textrm{se}1}}_{X'B\textrm{rev}}$ is the reverse of $\Pi^{B_{\textrm{se}1}}_{X'B\textrm{rev}}$. As is easily verified,\footnote{$\Pi=(\Pi_{XB\textrm{rev}}, - \Pi_{XC\textrm{rev}}, - \Pi_{X'B\textrm{rev}},  \Pi_{X'C\textrm{rev}})=X_1X'_2B_{\textrm{se}1}C_{\textrm{se}2} \xrightarrow{\rm wrev} X_2X'_2B_{\textrm{se}2}C_{\textrm{se}2} \xrightarrow{\rm wrev} X_1X'_2B_{\textrm{se}2}C_{\textrm{se}1}
\xrightarrow{\rm wrev} X_1X'_1B_{\textrm{se}1}C_{\textrm{se}1}
\xrightarrow{\rm wrev} X_1X'_2B_{\textrm{se}1}C_{\textrm{se}3}$. } the final state of the composite system $XX'BC$, after process $\Pi$, is $X_1 X'_2 B_{\textrm{se}1}C_{\textrm{se}3}$. Therefore, $\Pi$ is a reversible weight process for $C$ in which the regions of space occupied by the constituents of $C$ have no net change. If the energy of $C$ in state $C_{\textrm{se}3}$ were lower than that in the initial state $C_{\textrm{se}2}$, then $\Pi$ would violate Theorem \ref{Theorem1}. If the energy of $C$ in state $C_{\textrm{se}3}$ were higher than that in the initial state $C_{\textrm{se}2}$, then the reverse of $\Pi$ would violate Theorem \ref{Theorem1}. Therefore, the energy of $C$ in state $C_{\textrm{se}3}$ must coincide with the energy of $C$ in state $C_{\textrm{se}2}$, \textit{i.e.}, on account of Postulate \ref{Postulate2} and Lemma \ref{Lemma1}, the state $C_{\textrm{se}3}$ must coincide with $C_{\textrm{se}2}$.}
\end{lemma}

\begin{lemma}\label{LemmadE}
 For a given pair of systems, $B$ and $C$, consider an arbitrary pair of stable equilibrium states $(B_{\textrm{se}1}, C_{\textrm{se}1})$ 
 and the set of processes which defines the function  $E^C=f^{B\rightarrow C}_{11}(E^B)$ according to Lemma \ref{Lemma3}. Select another arbitrary stable equilibrium state  $B_{\textrm{se}2}$ of system $B$ and let $C_{\textrm{se}2}$ be the stable equilibrium state of system $C$ such that $E^C_{\textrm{se}2} = f^{B\rightarrow C}_{11}(E^B_{\textrm{se}2})$. Denote by $E^C=f^{B\rightarrow C}_{22}(E^B)$ the function defined by the set of reversible processes $\{(\Pi^{B_{\textrm{se}2}}_{XB\textrm{rev}}; \Pi^{C_{\textrm{se}2}}_{XC\textrm{rev}})\}$ according to Lemma \ref{Lemma3}. Then we have the identity
   \begin{equation}\label{new-equality44}
f^{B\rightarrow C}_{11}(E^B) =  f^{B\rightarrow C}_{22}(E^B) \mbox{ for every } E^B \ .
\end{equation}
 \dimostrazione{Lemma}{LemmadE}{Consider an arbitrary stable equilibrium state  of system $B$ with energy $E^B$ and denote it by $B_{\textrm{se}3}$, i.e. $E^B_{\textrm{se}3}=E^B$, and let $C_{\textrm{se}3}$ be the stable equilibrium state of system $C$ such that $E^C_{\textrm{se}3} = f^{B\rightarrow C}_{22}(E^B)$. Then, the pair of composite processes
$( X_1B_{\textrm{se}1} \xrightarrow{\rm wrev} X_2B_{\textrm{se}2} \xrightarrow{\rm wrev} X_3B_{\textrm{se}3} \;, \; X_1C_{\textrm{se}1} \xrightarrow{\rm wrev} X_2C_{\textrm{se}2} \xrightarrow{\rm wrev} X_3C_{\textrm{se}3})$ exists because $E^C_{\textrm{se}2} = f^{B\rightarrow C}_{11}(E^B_{\textrm{se}2})$ and $E^C_{\textrm{se}3} = f^{B\rightarrow C}_{22}(E^B_{\textrm{se}3})$, and it clearly belongs to the set of pairs of processes which defines the function $E^C=f^{B\rightarrow C}_{11}(E^B)$, therefore, $E^C_{\textrm{se}3} = f^{B\rightarrow C}_{11}(E^B_{\textrm{se}3})$.}
\end{lemma}


\begin{corollary}\label{Corollaryb}
The function $f^{B\rightarrow C}_{11}( E^B)$ defined through the set of pairs of processes $\{(\Pi^{B_{\textrm{se}1}}_{XB\textrm{rev}}; \Pi^{C_{\textrm{se}1}}_{XC\textrm{rev}})\}$ is strictly increasing. \dimostrazione{Corollary}{Corollaryb}{Consider the pairs of stable equilibrium states $(B_{\textrm{se}2}, C_{\textrm{se}2})$ and $(B_{\textrm{se}3}, C_{\textrm{se}3})$, such that $E^C_{\textrm{se}2} = f^{B\rightarrow C}_{11}(E^B_{\textrm{se}2})$, $E^C_{\textrm{se}3} = f^{B\rightarrow C}_{11}(E^B_{\textrm{se}3})$, and $E^B_{\textrm{se}3} > E^B_{\textrm{se}2}$. We will prove that $E^C_{\textrm{se}3} > E^C_{\textrm{se}2}$, \textit{i.e.}, $f^{B\rightarrow C}_{11}(E^B_{\textrm{se}3}) > f^{B\rightarrow C}_{11}(E^B_{\textrm{se}2})$.\\
Consider the pair of composite processes $( X_2B_{\textrm{se}2} \xrightarrow{\rm wrev} X_1B_{\textrm{se}1} \xrightarrow{\rm wrev} X_3B_{\textrm{se}3} \;, \; X_2C_{\textrm{se}2} \xrightarrow{\rm wrev} X_1C_{\textrm{se}1} \xrightarrow{\rm wrev} X_3C_{\textrm{se}3})$, which exists because $E^C_{\textrm{se}2} = f^{B\rightarrow C}_{11}(E^B_{\textrm{se}2})$ and $E^C_{\textrm{se}3} = f^{B\rightarrow C}_{11}(E^B_{\textrm{se}3})$. In this pair of processes, the energy change of $B$, $E^B_{\textrm{se}3} - E^B_{\textrm{se}2}$, is positive. On account of Theorem \ref{Theorem3}, also the energy change of $C$ must be positive, i.e., $E^C_{\textrm{se}3} > E^C_{\textrm{se}2}$.}
\end{corollary}


\begin{comment} Since the function $f^{B\rightarrow C}_{11}( E^B)$ is strictly increasing, it is invertible. The inverse of the function
\begin{equation}\label{function-direct}
E^C = f^{B\rightarrow C}_{11}( E^B)\;\; ,
\end{equation}
will be denoted by
\begin{equation}\label{inverse-function}
E^B = f^{C\rightarrow B}_{11}( E^C)\;\; .
\end{equation}
The domain of function $f^{B\rightarrow C}_{11}$ is the set of all the energy values of system $B$ compatible with the regions of space occupied by the constituents of $B$ in state $B_{\textrm{se}1}$. The domain of function $f^{C\rightarrow B}_{11}$ is the set of all the energy values of system $C$ compatible with the regions of space occupied by the constituents of $C$ in state $C_{\textrm{se}1}$.\end{comment}

\begin{lemma}\label{Lemma4}
Consider three systems $B$, $C$, and $R$,  three stable equilibrium states $B_{\textrm{se}1}$, $C_{\textrm{se}1}$, and  $R_{\textrm{se}1}$, and the functions $E^R = f^{B\rightarrow R}_{11}( E^B)$, $E^C = f^{R\rightarrow C}_{11}( E^R)$, and $E^C = f^{B\rightarrow C}_{11}( E^B)$ defined as in Lemma \ref{Lemma3}. Then,
\begin{equation}\label{function-function}
f^{B\rightarrow C}_{11}( E^B)=f^{R\rightarrow C}_{11}( f^{B\rightarrow R}_{11}( E^B))\;\; .
\end{equation}
 \dimostrazione{Lemma}{Lemma4}{Consider an auxiliary system $X$, the pair of states $(X_1,X_2)$, and the three processes  $\Pi^{B_{\textrm{se}1}}_{XB\textrm{rev}}$, $\Pi^{C_{\textrm{se}1}}_{XC\textrm{rev}}$, $\Pi^{R_{\textrm{se}1}}_{XR\textrm{rev}}$, respectively defined as follows: $\Pi^{B_{\textrm{se}1}}_{XB\textrm{rev}}$ is a reversible weight process for $XB$ with initial and final states $X_1$ and $X_2$ for $X$, and initial state $B_{\textrm{se}1}$ for $B$;  $\Pi^{C_{\textrm{se}1}}_{XC\textrm{rev}}$ is a reversible weight process for $XC$ with initial and final states $X_1$ and $X_2$ for $X$, and initial state $C_{\textrm{se}1}$ for $C$; $\Pi^{R_{\textrm{se}1}}_{XR\textrm{rev}}$ is a reversible weight process for $XR$ with initial and final states $X_1$ and $X_2$ for $X$, and initial state $R_{\textrm{se}1}$ for $R$. Let us denote by $E^B_{\textrm{se}2}$, $E^C_{\textrm{se}2}$, $E^R_{\textrm{se}2}$ the energy of the final states of $B$, $C$ and $R$, respectively.
The pair of processes $(\Pi^{B_{\textrm{se}1}}_{XB\textrm{rev}},\Pi^{R_{\textrm{se}1}}_{XR\textrm{rev}})$ belongs to the set of processes $\{(\Pi^{B_{\textrm{se}1}}_{XB\textrm{rev}},\Pi^{R_{\textrm{se}1}}_{XR\textrm{rev}})\}$ that defines according to Lemma \ref{Lemma3} the function $E^R = f^{B\rightarrow R}_{11}( E^B)$, therefore,
\begin{equation}\label{ffBR}
E^R_{\textrm{se}2} = f^{B\rightarrow R}_{11}( E^B_{\textrm{se}2}) \ .
\end{equation}
The pair of processes $(\Pi^{R_{\textrm{se}1}}_{XR\textrm{rev}},\Pi^{C_{\textrm{se}1}}_{XC\textrm{rev}})$ belongs to the set of processes $\{(\Pi^{R_{\textrm{se}1}}_{XR\textrm{rev}},\Pi^{C_{\textrm{se}1}}_{XC\textrm{rev}})\}$ that defines according to Lemma \ref{Lemma3} the function $E^C = f^{R\rightarrow C}_{11}( E^R)$, therefore,
\begin{equation}\label{ffRC}
E^C_{\textrm{se}2} = f^{R\rightarrow C}_{11}( E^R_{\textrm{se}2}) \ .
\end{equation}
The pair of processes $(\Pi^{B_{\textrm{se}1}}_{XB\textrm{rev}},\Pi^{C_{\textrm{se}1}}_{XC\textrm{rev}})$ belongs to the set of processes $\{(\Pi^{B_{\textrm{se}1}}_{XB\textrm{rev}},\Pi^{C_{\textrm{se}1}}_{XC\textrm{rev}})\}$ that defines according to Lemma \ref{Lemma3} the function $E^C = f^{B\rightarrow C}_{11}( E^B)$, therefore,
\begin{equation}\label{ffBC}
E^C_{\textrm{se}2} = f^{B\rightarrow C}_{11}( E^B_{\textrm{se}2}) \ .
\end{equation}
From (\ref{ffBR}) and (\ref{ffRC}) it follows that
\begin{equation}\label{fff}
E^C_{\textrm{se}2} =f^{R\rightarrow C}_{11}( f^{B\rightarrow R}_{11}( E^B_{\textrm{se}2}))\;\; .
\end{equation}
Comparing (\ref{fff}) and (\ref{ffBC}) we find
\begin{equation}\label{fff2}
f^{B\rightarrow C}_{11}( E^B_{\textrm{se}2}) =f^{R\rightarrow C}_{11}( f^{B\rightarrow R}_{11}( E^B_{\textrm{se}2}))\;\; .
\end{equation}
Equation (\ref{function-function}) follows immediately from (\ref{fff2}) by repeating the above for all possible choices of the pair of states $(X_1,X_2)$.}
\end{lemma}

\begin{assumption}\label{Assumption4} The function $f^{B\rightarrow C}_{11}( E^B)$ defined through the set of pairs of processes $\{(\Pi^{B_{\textrm{se}1}}_{XB\textrm{rev}}; \Pi^{C_{\textrm{se}1}}_{XC\textrm{rev}})\}$ is differentiable in $E^B_{\textrm{se}1}$; in symbols
\begin{equation}\label{limit}
\lim_{E^B \rightarrow E^B_{\textrm{se}1}} \frac{f^{B\rightarrow C}_{11}( E^B) - f^{B\rightarrow C}_{11}( E^B_{\textrm{se}1})}{E^B - E^B_{\textrm{se}1}} = \frac{\textrm{d}f^{B\rightarrow C}_{11}}{\textrm{d}E^B}\bigg|_{E^B_{\textrm{se}1}} \;\; .
\end{equation}
\end{assumption}

\begin{corollary}\label{Corollaryc}
The inverse function $E^B = f^{C\rightarrow B}_{11}( E^C)$ is differentiable in
$E^C_{\textrm{se}1}$, moreover if  $\textrm{d}f^{B\rightarrow C}_{11}/\textrm{d}E^B\big|_{E^B_{\textrm{se}1}}\ne 0$ then
\begin{equation}\label{derivative-inverse}
\frac{\textrm{d}f^{C\rightarrow B}_{11}}{\textrm{d}E^C}\bigg|_{E^C_{\textrm{se}1}} = \frac{1}{\displaystyle\frac{\textrm{d}f^{B\rightarrow C}_{11}}{\textrm{d}E^B}\bigg|_{E^B_{\textrm{se}1}}}\;\; .
\end{equation}
 \dimostrazione{Corollary}{Corollaryc}{Since Assumption \ref{Assumption4} holds for any pair of systems, by exchanging $B$ with $C$ it implies that also the function $f^{C\rightarrow B}_{11}( E^C)$ is differentiable. Equation (\ref{derivative-inverse}) follows from the theorem on the derivative of the inverse function.}
\end{corollary}


\subsection*{Temperature of a stable equilibrium state.}
Let $R$ be a \textit{reference system}, and let $R_{\rm se1}$ be a
\textit{reference stable equilibrium state} of $R$.
Both $R$ and $R_{\rm se1}$ are fixed once and for all, and a positive real number, $T^R_{\rm se1}$, chosen arbitrarily, is associated with
$R_{\rm se1}$ and called \textit{temperature} of $R_{\rm se1}$. Let $B$ be any system, and $B_{\textrm{se}1}$ any stable equilibrium state of $B$.\\
Let us consider the set of pairs of processes $\{(\Pi^{R_{\textrm{se}1}}_{XR\textrm{rev}}; \Pi^{B_{\textrm{se}1}}_{XB\textrm{rev}})\}$, where $\Pi^{R_{\textrm{se}1}}_{XR\textrm{rev}}$ is any reversible weight processes for $XR$ standard with respect $R$ and with initial state $R_{\textrm{se}1}$, $\Pi^{B_{\textrm{se}1}}_{XB\textrm{rev}}$ is any reversible weight processes for $XB$ standard with respect $B$ and with initial state $B_{\textrm{se}1}$, and $X$ is a system which can be chosen and changed arbitrarily, as well as the initial state of $X$. On account of Lemma \ref{Lemma3} and of Assumption \ref{Assumption4}, the set of pairs of processes $\{(\Pi^{R_{\textrm{se}1}}_{XR\textrm{rev}}; \Pi^{B_{\textrm{se}1}}_{XB\textrm{rev}})\}$ defines a single valued and invertible function $f^{R\rightarrow B}_{11}( E^R)$, from the energy values of the stable equilibrium states of $R$ with the same regions of space  as $R_{\textrm{se}1}$ to the energy values of the stable equilibrium states of $B$ with the same regions of space  as $B_{\textrm{se}1}$, which is differentiable in $E^R_{\textrm{se}1}$. We define as \textit{temperature} of system $B$ in the stable equilibrium state $B_{\rm se1}$ the quantity
\begin{equation}\label{temperature}
\!\frac{T^B_{\rm se1}}{T^R_{\rm se1}} \!=\!\!\! \lim_{E^R \rightarrow E^R_{\textrm{se}1}} \!\!\!\! \frac{f^{R\rightarrow B}_{11}\!( E^R) - f^{R\rightarrow B}_{11}\!( E^R_{\textrm{se}1})}{E^R - E^R_{\textrm{se}1}} \!=\!
\frac{\textrm{d}f^{R\rightarrow B}_{11}}{\textrm{d}E^R}\bigg|_{E^R_{\textrm{se}1}} \! .
\end{equation}
On account of Corollary \ref{Corollaryb}, $T^B_{\rm se1}$ is non-negative. Since $R$ and $R_{\textrm{se}1}$ have been fixed once and for all, the temperature is a property of $B$, defined for all the stable equilibrium states of $B$. Clearly, the property temperature is defined by Eq.\ (\ref{temperature}) only with respect to the chosen reference state $R_{\rm se1}$ of the reference system $R$ and up to the arbitrary multiplicative constant $T^R_{\rm se1}$.

\begin{corollary}\label{CorollaryTfunction}
The temperature of the stable equilibrium states of any system $B$ is a function of its energy $E^B$ and the region of space $\RB{}$  it occupies, i.e.,
\begin{equation}\label{TfunctionER}
T^B=T^B(E^B;\RB{}) \ ,
\end{equation}
provided the reference state $R_{\rm se1}$ of the reference system $R$ and the arbitrary multiplicative constant $T^R_{\rm se1}$ that are necessary for the definition of $T^B$ according to Eq.\ (\ref{temperature}) have been chosen once and for all. \dimostrazione{Corollary}{CorollaryTfunction}{The conclusion is a direct consequence of Postulate \ref{Postulate2}, Lemma \ref{Lemma1} and definition (\ref{temperature}).}
\end{corollary}


\subsection*{Choice of the reference system and of the
reference stable equilibrium state.} In the macroscopic domain, the following choice of $R$
and of $R_{\rm se1}$ is currently employed, because it can be
easily reproduced in any laboratory. The reference system $R$ is
composed of a sufficient number of moles of pure water and its
reference stable equilibrium state $R_{\rm se1}$ is any of the
stable equilibrium states of $R$ in which ice, liquid water, and
water vapor coexist. This choice is convenient because, up to the
measurement accuracy available today, the value of the limit in
Eq.\ (\ref{temperature}) is practically independent of both the
number of moles in system $R$ and the particular choice of the
reference state $R_{\textrm{se}1}$, as long as it belongs to the
set of triple-point states. For all practical purposes, Assumption
\ref{Assumption3} is satisfied for all macroscopic systems $A$ by taking as system $R$ a sufficiently large number of moles of water and as state $R_{\textrm{se}1}$ a suitable triple-point state. With this selection for the reference stable equilibrium state, we obtain the S.I. thermodynamic
temperature, with unit called \textit{kelvin}, by setting
$T^R_{\textrm{se}1}=273.16$ K.\\
In the microscopic field, it could be convenient to choose as a reference system $R^\mu$ a few-particle monoatomic gas and as a reference state of $R^\mu$ a stable equilibrium state $R^\mu_{\textrm{se}1}$ such that $R^\mu$ is in mutual stable equilibrium with water at the triple point. Thus, the temperature of $R^\mu$ in state $R^\mu_{\textrm{se}1}$ would coincide with that of $R$ in state $R_{\rm se1}$, as we  prove in Theorem \ref{Theorem13}. Note that, by the next theorem (Theorem \ref{Theorem4}), we  prove that the ratio of two temperatures can be measured directly and is independent of the choice of the reference system and of the reference stable equilibrium state. Hence, any system in any stable equilibrium state such that the temperature of the system is known can be used as a new reference system in a reference stable equilibrium state, without inconsistencies.

\begin{theorem}\label{Theorem4}
Let $B_{\rm se1}$ be any stable equilibrium state of a system $B$ and let $C_{\rm se1}$ be any stable equilibrium state of a system $C$, both with a non vanishing temperature. Then, the ratio of the temperatures of $B_{\rm se1}$ and $C_{\rm se1}$, as defined via Eq.\ (\ref{temperature}),
is independent of the choice of the reference system $R$ and of the reference stable equilibrium state $R_{\rm se1}$, and can be measured directly by the following procedure.\\
Consider the set of pairs of processes $\{(\Pi^{B_{\textrm{se}1}}_{XB\textrm{rev}}; \Pi^{C_{\textrm{se}1}}_{XC\textrm{rev}})\}$, where $\Pi^{B_{\textrm{se}1}}_{XB\textrm{rev}}$ is any reversible weight processes for $XB$ standard with respect $B$ and with initial state $B_{\textrm{se}1}$, $\Pi^{C_{\textrm{se}1}}_{XC\textrm{rev}}$ is any reversible weight processes for $XC$ standard with respect $C$, with initial state $C_{\textrm{se}1}$ and with the same initial and final state of $X$ as $\Pi^{B_{\textrm{se}1}}_{XB\textrm{rev}}$, and $X$ is a system which can be chosen and changed arbitrarily, as well as the initial state of $X$. On account of Lemma \ref{Lemma3} the set of pairs of processes $\{(\Pi^{B_{\textrm{se}1}}_{XB\textrm{rev}}; \Pi^{C_{\textrm{se}1}}_{XC\textrm{rev}})\}$ defines a single valued and invertible function $f^{B\rightarrow C}_{11}( E^B)$, which is differentiable in $E^B_{\textrm{se}1}$. The ratio of the temperatures $T^C_{\textrm{se}1}$ and $T^B_{\textrm{se}1}$ is given by
\begin{equation}\label{temperature-direct}
\!\frac{T^C_{\rm se1}}{T^B_{\rm se1}} \!=\!\!\! \lim_{E^B \rightarrow E^B_{\textrm{se}1}} \!\!\!\! \frac{f^{B\rightarrow C}_{11}\!( E^B) - f^{B\rightarrow C}_{11}\!( E^B_{\textrm{se}1})}{E^B - E^B_{\textrm{se}1}} \!=\!
\frac{\textrm{d}f^{B\rightarrow C}_{11}}{\textrm{d}E^B}\bigg|_{E^B_{\textrm{se}1}} \! .\!\!
\end{equation}
 \dimostrazione{Theorem}{Theorem4}{By applying to Eq.\ (\ref{function-function}) the theorem on the derivative of a composite function, one obtains
\begin{equation}\label{derivative-composite}
\frac{\textrm{d} f^{B\rightarrow C}_{11}}{\textrm{d} E^B}\bigg|_{E^B_{\textrm{se}1}} \!\!= \frac{\textrm{d} f^{R\rightarrow C}_{11}}{\textrm{d}  E^R}\bigg|_{E^R=f^{B\rightarrow R}_{11}( E^B_{\textrm{se}1})} \; \frac{\textrm{d} f^{B\rightarrow R}_{11}}{\textrm{d} E^B}\bigg|_{E^B_{\textrm{se}1}} .
\end{equation}
On account of Eq.\ (\ref{temperature}), the first derivative at the right hand side of Eq.\ (\ref{derivative-composite}) can be rewritten as
\begin{equation}\label{derivative-composite-first}
 \frac{\textrm{d} f^{R\rightarrow C}_{11}}{\textrm{d} E^R}\bigg|_{E^R_{\textrm{se}1}} = \frac{T^C_{\rm se1}}{T^R_{\rm se1}} \;\; .
\end{equation}
By applying Eqs.\ (\ref{derivative-inverse}) and (\ref{temperature}), the second derivative at the right hand side of Eq.\ (\ref{derivative-composite}) can be rewritten as
\begin{equation}\label{derivative-composite-second}
\frac{\textrm{d} f^{B\rightarrow R}_{11}}{\textrm{d} E^B}\bigg|_{E^B_{\textrm{se}1}} = \frac{1}{\displaystyle \frac{\textrm{d} f^{R\rightarrow B}_{11}}{\textrm{d} E^R}\bigg|_{E^R_{\textrm{se}1}}} = \frac{1}{\displaystyle \frac{T^B_{\rm se1}}{T^R_{\rm se1}}} = \frac{T^R_{\rm se1}}{T^B_{\rm se1}} \;\; .
\end{equation}
By combining Eqs.\ (\ref{derivative-composite}), (\ref{derivative-composite-first}) and (\ref{derivative-composite-second}) we obtain Eq.\ (\ref{temperature-direct}).} Theorem \ref{Theorem4} completes the definition of  temperature of a stable equilibrium state.
\end{theorem}


\section{Definition of entropy for any state}

\begin{corollary}\label{Corollarydf22}
Consider a pair of stable equilibrium states $(B_{\textrm{se}1}, C_{\textrm{se}1})$ and the set of processes which defines the function  $E^C=f^{B\rightarrow C}_{11}(E^B)$ according to Lemma \ref{Lemma3}. Then, for every pair of stable equilibrium states of $B$ and $C$ determined by the same regions of space $\RB{}$ and $\RC{}$ as $B_{\textrm{se}1}$ and $C_{\textrm{se}1}$, respectively, and by the energy values $E^B$ and $E^C=f^{B\rightarrow C}_{11}(E^B)$,
\begin{equation}\label{new-equality22}
\frac{T^C\!(E^C\!\!=\! f^{B\rightarrow C}_{11}\!(E^B);\RC{})}{T^B\!(E^B;\RB{})} = \left.\frac{\textrm{d}f^{B\rightarrow C}_{11}\!(E^B)}{\textrm{d}E^B}\right|_{E^B}\ .
\end{equation}
 \dimostrazione{Corollary}{Corollarydf22}{For the fixed regions of space $\RB{}$, consider the set of stable equilibrium states  of system $B$ defined by varying the energy $E^B$. Select a value of energy $E^B$ and denote the corresponding  state in this set by $B_{\textrm{se}2}$, i.e., $E^B_{\textrm{se}2}=E^B$. Consider the pair of stable equilibrium states $(B_{\textrm{se}2}, C_{\textrm{se}2})$, where $C_{\textrm{se}2}$ is such that $E^C_{\textrm{se}2} = f^{B\rightarrow C}_{11}(E^B_{\textrm{se}2})$ and let $E^C=f^{B\rightarrow C}_{22}(E^B)$ be the function defined according to Lemma \ref{Lemma3}. Then, we have
\begin{equation}\label{new-equality33}
\frac{T^C\!(E^C_{\textrm{se}2};\RC{})}{T^B\!(E^B_{\textrm{se}2};\RB{})} \!=\!\frac{T^C_{\textrm{se}2}}{T^B_{\textrm{se}2}} \!=\! \left.\frac{\textrm{d}f^{B\rightarrow C}_{22}\!(E^B)}{\textrm{d}E^B}\right|_{E^B_{\textrm{se}2}} \!\!\!=\! \left.\frac{\textrm{d}f^{B\rightarrow C}_{11}\!(E^B)}{\textrm{d}E^B}\right|_{E^B_{\textrm{se}2}} \!,
\end{equation}
where the first equality obtains from Eq.\ (\ref{TfunctionER}), the second equality  from Eq.\ (\ref{temperature-direct}) applied to $f^{B\rightarrow C}_{22}(E^B)$, and the third  from  Eq.\ (\ref{new-equality44}).
Recalling that $E^B_{\textrm{se}2}=E^B$, that $E^B$ can be varied arbitrarily, and that $E^C_{\textrm{se}2}=f^{B\rightarrow C}_{11}(E^B)$,  Eq.\ (\ref{new-equality33}) yields Eq.\ (\ref{new-equality22}).}
\end{corollary}


\begin{assumption}\label{Assumption5} For every system $B$ and every choice of the regions of space $\RB{}$ occupied by the constituents of $B$, the temperature of $B$ is a continuous function of the energy of $B$ and is vanishing only in the stable equilibrium state with lowest energy, which is called \textit{ground state}.
\end{assumption}

\begin{lemma}\label{Lemma5}
For every pair of stable equilibrium states $B_{\textrm{se}1}$ and $B_{\textrm{se}2}$ of a system $B$, with a non vanishing temperature and with the same regions of space $\RB{}$ occupied by the constituents of $B$, the integral
\begin{equation}\label{integral-A}
\int_{E^B_{\textrm{se}1}}^{E^B_{\textrm{se}2}}
\frac{1}{T^B(E^B;\RB{})} \; \textrm{d}E^B \;\; ,
\end{equation}
 has a finite value and the same sign as $E^B_{\textrm{se}2} - E^B_{\textrm{se}1}$. \dimostrazione{Lemma}{Lemma5}{Since both $E^B_{\textrm{se}1}$ and $E^B_{\textrm{se}2}$ are greater than the lowest energy value for the given regions of space $\RB{}$, on account of Assumption \ref{Assumption5} the function $1 /\,T^B(E^B;\RB{})$ is defined and continuous in the whole interval. Therefore the integral in Eq.\ (\ref{integral-A}) exists. Moreover, on account of Corollary \ref{Corollaryb}, the function $1 /\,T^A(E;\RA{})$ has positive values. Therefore, if $E^A_{\textrm{se}2} > E^A_{\textrm{se}1}$ the integral in Eq.\ (\ref{integral-A}) has a positive value; if $E^A_{\textrm{se}2} < E^A_{\textrm{se}1}$ the integral in Eq.\ (\ref{integral-A}) has a negative value.}
 \end{lemma}


\begin{theorem}\label{Theorem5}
Consider an arbitrarily chosen pair of states  $(A_1, A_2)$ of a system $A$, such that $A$ is separable and uncorrelated from its environment, another system $B$ in the environment of $A$ and a
reversible weight process $\Pi^{B_{\textrm{se}1}}_{AB\rm rev}$ for $AB$ in which $A$ goes from $A_1$ to $A_2$, standard with respect to $B$ and with initial state $B_{\textrm{se}1}$, chosen so that the temperature of $B$ is non vanishing both for $B_{\textrm{se}1}$ and for the final state $B_{\textrm{se}2}$. Denote by $\RB{}$ the regions of space occupied by the constituents of $B$ in its end states $B_{\textrm{se}1}$ and
$B_{\textrm{se}2}$. Then the value of the integral
\begin{equation}\label{integral}
\int_{E^B_{\textrm{se}1}}^{E^B_{\textrm{se}2}}
\frac{1}{T^B(E^B;\RB{})} \; \textrm{d}E^B \;\; ,
\end{equation}
depends only on the
pair of states $(A_1, A_2)$ of system $A$ and is independent of the choice of
system $B$, of the initial stable equilibrium state
$B_{\textrm{se}1}$, and of the details of the reversible
weight process for $AB$, standard with respect to $B$.  \dimostrazione{Theorem}{Theorem5}{On account of Theorem \ref{Theorem2}, once the initial state $B_{\textrm{se}1}$ has been chosen, the final state $B_{\textrm{se}2}$ is determined uniquely. Therefore, the value of the integral in Eq.\ (\ref{integral}) can depend, at most, on the pair of states $(A_1, A_2)$ and on the choice of system $B$ and of its initial state $B_{\textrm{se}1}$.
Consider another system $C$ and a reversible weight process $\Pi^{C_{\textrm{se}1}}_{AC\rm rev}$ for $AC$ in which $A$  goes again from $A_1$ to $A_2$, standard with respect to $C$ and with an initial state $C_{\textrm{se}1}$ chosen arbitrarily, provided that the temperature of $C$ is non vanishing both for $C_{\textrm{se}1}$ and for the final state $C_{\textrm{se}2}$. We will prove that the integral
\begin{equation}\label{integral-C}
\int_{E^C_{\textrm{se}1}}^{E^C_{\textrm{se}2}}
\frac{1}{T^C(E^C;\RC{})} \; \textrm{d}E^C \;\;
\end{equation}
has the same value as the integral in Eq.\ (\ref{integral}), implying that such value is independent of the choice of system $B$ and of the initial state $B_{\textrm{se}1}$, and, therefore, it depends only on the pair of states $(A_1, A_2)$.\\
The set of pairs of processes $\{(\Pi^{B_{\textrm{se}1}}_{AB\rm rev},\Pi^{C_{\textrm{se}1}}_{AC\rm rev})\}$ such that the energy of the final state of $B$ is in the range $E^B_{\textrm{se}1}\le E^B \le E^B_{\textrm{se}2} $  belongs to the set defined in Lemma \ref{Lemma3}, so that $E^C=f_{11}^{B\rightarrow C} (E^B)$ and, since this function is invertible (Lemma \ref{Lemma3}), $E^B=f_{11}^{C\rightarrow B} (E^C)$ so that, in particular, $E^B_{\textrm{se}1}=f_{11}^{C\rightarrow B} (E^C_{\textrm{se}1})$ and  $E^B_{\textrm{se}2}=f_{11}^{C\rightarrow B} (E^C_{\textrm{se}2})$. Now, consider the change of integration variable in the definite integral (\ref{integral-C}) from $E^C=f_{11}^{B\rightarrow C} (E^B)$ to $E^B$. By virtue of Eq.\ (\ref{new-equality22}) (Corollary \ref{Corollarydf22}) we have
\begin{equation}\label{new-equality1}
    \textrm{d}E^C\!=\!\frac{\textrm{d}f^{B\rightarrow C}_{11}(E^B)}{\textrm{d}E^B}\;\textrm{d}E^B \!=\! \frac{T^C(f_{11}^{B\rightarrow C} (E^B);\RC{})}{T^B(E^B;\RB{})}\;\textrm{d}E^B  .
\end{equation}
Thus, the integral in Eq.\ (\ref{integral-C}) can be rewritten as follows
\begin{eqnarray}\label{integral-equality}
&&\!\!\!\!\!\!\!\!\!\!\!\int_{f_{11}^{C\rightarrow B} (E^C_{\textrm{se}1})}^{f_{11}^{C\rightarrow B} (E^C_{\textrm{se}2})}
\frac{1}{T^C(f_{11}^{B\rightarrow C} (E^B);\RC{})} \frac{T^C(f_{11}^{B\rightarrow C} (E^B);\RC{})}{T^B(E^B;\RB{})}\;\textrm{d}E^B \nonumber\\ &&\!\!\!\!\!\!\!\!\!\!\!= \int_{E^B_{\textrm{se}1}}^{E^B_{\textrm{se}2}}
\frac{1}{T^B(E;\RB{})} \; \textrm{d}E^B  \end{eqnarray}}
\end{theorem}

\subsection*{Definition of (thermodynamic) entropy, proof that
it is a property.} Let ($A_1$,$A_2$) be any pair of states of a
system $A$, such that $A$ is separable and uncorrelated from its environment,
and let $B$ be any other system placed in the
environment of $A$. We call \textit{entropy difference} between
$A_2$ and $A_1$ the quantity
\begin{equation}\label{entropy}
S^A_2 - S^A_1 = - \int_{E^B_{\textrm{se}1}}^{E^B_{\textrm{se}2}}
\frac{1}{T^B(E^B;\RB{})} \; \textrm{d}E^B \;\; ,
\end{equation}
where $B_{\textrm{se}1}$ and $B_{\textrm{se}2}$ are the initial
and the final state of $B$ in any reversible weight process for
$AB$ from $A_1$ to $A_2$, standard with respect to $B$, $\RB{}$ is
the set of regions of space occupied by the constituents of $B$ in
the states $B_{\textrm{se}1}$ and $B_{\textrm{se}2}$, and $T^B$ is the
temperature of $B$. The initial state $B_{\textrm{se}1}$ is chosen so that both $T^B_{\textrm{se}1}$ and $T^B_{\textrm{se}2}$ are non vanishing.
On account of Theorem \ref{Theorem5}, the right hand side
of Eq.\ (\ref{entropy}) is determined uniquely by states $A_1$ and
$A_2$.\\ Let $A_0$ be a reference state of $A$, to which we assign
an arbitrarily chosen value of entropy $S^A_0$. Then, the value of
the entropy of $A$ in any other state $A_1$ of $A$ such that $A$ is separable and uncorrelated from its environment is determined uniquely by the equation
\begin{equation}\label{entropyabs}
S^A_1 = S^A_0  - \int_{E^B_{\textrm{se}1}}^{E^B_{\textrm{se}2}}
\frac{1}{T^B(E;\RB{})} \; \textrm{d}E^B \;\; ,
\end{equation}
where $B_{\textrm{se}1}$ and $B_{\textrm{se}2}$ are the initial
and the final state of $B$ in any reversible weight process for
$AB$ from $A_0$ to $A_1$, standard with respect to $B$, $T^B_{\textrm{se}1}$ and $T^B_{\textrm{se}2}$ are non vanishing, and the
other symbols have the same meaning as in Eq.\ (\ref{entropy}).
Such a process exists for every state $A_1$ such that $A$ is separable and uncorrelated from its environment, in a set of states where Assumption \ref{Assumption3} holds.

\section{Principle of entropy non-decrease, additivity of entropy, maximum entropy principle}

Based on the above construction, in this section we obtain some of the main standard theorems about entropy and entropy change.

\begin{lemma}\label{Lemma7}  Let ($A_1$, $A_2$) be any pair of
states of a system $A$ such that $A$ is separable and uncorrelated from its environment, and let $B$ be any other system placed in the environment of $A$. Let $\Pi_{AB\rm irr}$ be any irreversible weight process for $AB$, standard with respect to $B$, from $A_1$ to $A_2$, and let $B_{\textrm{se}1}$ and $B_{\textrm{se}2\textrm{irr}}$ be the end states of $B$ in the process. Then
\begin{equation}\label{inequality}
- \int_{E^B_{\textrm{se}1}}^{E^B_{\textrm{se}2\textrm{irr}}}
\frac{1}{T^B(E^B;\RB{})} \; \textrm{d}E^B < S^A_2 - S^A_1 \;\; .
\end{equation}
 \dimostrazione{Lemma}{Lemma7}{Let $\Pi_{AB\rm rev}$ be any reversible weight process
for $AB$, standard with respect to $B$, from $A_1$ to $A_2$, with the same initial state $B_{\textrm{se}1}$ of $B$, and let $B_{\textrm{se}2\textrm{rev}}$ be the final state of $B$ in this process. On
account of Theorem \ref{Theorem2},
\begin{equation}\label{minimum-energychange}
E^B_{\textrm{se}2\textrm{rev}} < E^B_{\textrm{se}2\textrm{irr}} \;\; .
\end{equation}
Since $T^B$ is a positive function, from Eqs.\ (\ref{minimum-energychange})
and (\ref{entropy}) one obtains
\begin{equation}\label{inequality-two}
\!- \!\!\int_{E^B_{\textrm{se}1}}^{E^B_{\textrm{se}2\textrm{irr}}} \!
\frac{1}{T^B(E^B;\RB{})} \; \textrm{d}E^B \! <\!  - \!\! \int_{E^B_{\textrm{se}1}}^{E^B_{\textrm{se}2\textrm{rev}}} \!\!
\frac{1}{T^B(E^B;\RB{})} \; \textrm{d}E^B \!= \! S^A_2\! -\! S^A_1.
\end{equation}}
\end{lemma}


\begin{theorem}\label{Theorem6} \textbf{Principle of entropy non-decrease in weight processes}.
Let $(A_1, A_2)$ be a pair of  states of a system $A$ such that $A$ is separable and uncorrelated from its environment and let $A_1\xrightarrow{\rm w}A_2$ be any weight process for $A$ from $A_1$
to $A_2$. Then, the entropy difference $S^A_2 - S^A_1$ is equal to
zero if and only if the weight process is reversible; it is
strictly positive if and only if the weight process is
irreversible. \dimostrazione{Theorem}{Theorem6}{If  $A_1\xrightarrow{\rm w}A_2$ is
reversible, then it is a special case of a reversible
weight process for $AB$, standard with respect to $B$, in which the initial stable equilibrium
state of $B$ does not change. Therefore, $E^B_{\textrm{se}2\textrm{rev}} = E^B_{\textrm{se}1}$ and  Eq.\ (\ref{entropy}) yields
\begin{equation}\label{nondecrease-rev}
S^A_2 - S^A_1 = - \int_{E^B_{\textrm{se}1}}^{E^B_{\textrm{se}2\textrm{rev}}}
\frac{1}{T^B(E^B;\RB{})} \; \textrm{d}E^B = 0 \;\; .
\end{equation}
If  $A_1\xrightarrow{\rm w}A_2$ is irreversible, then it is a
special case of an irreversible weight process for $AB$, standard with respect to $B$,
in which the initial stable equilibrium state of $B$ does not
change. Therefore, $E^B_{\textrm{se}2\textrm{irr}} = E^B_{\textrm{se}1}$ and Eq.
(\ref{inequality})  yields
\begin{equation}\label{nondecrease-irr}
S^A_2 - S^A_1 > - \int_{E^B_{\textrm{se}1}}^{E^B_{\textrm{se}2\textrm{irr}}}
\frac{1}{T^B(E^B;\RB{})} \; \textrm{d}E^B = 0 \;\; .
\end{equation}
Moreover, if a weight process $A_1\xrightarrow{\rm w}A_2$ for
$A$ is such that $S^A_2 - S^A_1 = 0$, then the process must be
reversible, because we just proved that for any irreversible
weight process  $S^A_2 - S^A_1 > 0$; if a weight process $A_1\xrightarrow{\rm w}A_2$ for $A$ is such that $S^A_2 - S^A_1 > 0$, then
the process must be irreversible, because we just proved that for
any reversible weight process $S^A_2 - S^A_1 = 0$.}
\end{theorem}


\begin{theorem}\label{Theorem7} \textbf{ Additivity of entropy differences}.
Consider the pair of states $(C_1 = A_1B_1, C_2 = A_2 B_2)$ of the
composite system $C =A B$, such that $A$, $B$ and $C$ are separable and uncorrelated from their environment. Then,
\begin{equation}\label{entropyadditivity}
S^{AB}_{A_2 B_2} - S^{AB}_{A_1 B_1} = S^A_2 - S^A_1 + S^B_2 -
S^B_1 \;\; .
\end{equation}
 \dimostrazione{Theorem}{Theorem7}{Let us choose a system $D$ (with fixed
regions of space $\RD{}$) in the environment of $C$, and consider the processes  $\Pi_{AD\rm
rev}=A_1D_{\rm se1}\xrightarrow{\rm wrev}A_2D_{\rm se3rev}$ and
$\Pi_{BD\rm rev}=B_1D_{\rm se3rev}\xrightarrow{\rm wrev}B_2D_{\rm
se2rev}$. For process $\Pi_{AD\rm rev}$ Eq.\ (\ref{entropy}) implies that
\begin{equation}\label{Proof7-A}
S^A_2 - S^A_1 =- \int_{E^D_{\textrm{se}1}}^{E^D_{\textrm{se}3\textrm{rev}}}
\frac{1}{T^D(E^D;\RD{})} \; \textrm{d}E^D\ .\end{equation} For process
$\Pi_{BD\rm rev}$ Eq.\ (\ref{entropy}) implies that
\begin{equation}\label{Proof7-B}S^B_2 - S^B_1 =- \int_{E^D_{\textrm{se}3\textrm{rev}}}^{E^D_{\textrm{se}2\textrm{rev}}}
\frac{1}{T^D(E^D;\RD{})} \; \textrm{d}E^D\ .\end{equation} The composite process
$(\Pi_{AD\rm rev},\Pi_{BD\rm rev}) = A_1B_1D_{\rm
se1} \xrightarrow{\rm wrev} A_2B_1D_{\rm se3rev} \xrightarrow{\rm
wrev} A_2B_2D_{\rm se2rev}$ is a reversible weight process from
$C_1=A_1B_1$ to $C_2=A_2 B_2$ for $CD$, standard with respect to
$D$, in which the energy change of $D$ is the sum of its energy
changes in the constituent processes $\Pi_{AD\rm rev}$ and
$\Pi_{BD\rm rev}$. Therefore, Eq.\ (\ref{entropy}) implies that
\begin{equation}\label{Proof7-C}S^C_2 - S^C_1 =- \int_{E^D_{\textrm{se}1}}^{E^D_{\textrm{se}2\textrm{rev}}}
\frac{1}{T^D(E^D;\RD{})} \; \textrm{d}E^D\ .\end{equation} Subtracting Eqs.
(\ref{Proof7-A}) and  (\ref{Proof7-B}) from Eq.\ (\ref{Proof7-C})
yields Eq.\ (\ref{entropyadditivity}).}
\end{theorem}


\begin{comment} As a consequence of Theorem \ref{Theorem7}, if the
values of entropy are chosen so that they are additive in the
reference states, entropy results as an additive property.\end{comment}

\spazio \begin{theorem}\label{Theorem8}\textbf{Maximum entropy principle}. Consider a closed system $A$, and the set of all the states of $A$ with a given value $E^A_1$ of the energy, given regions of space $\RA{}$, and
such that $A$ is separable and uncorrelated from its environment.
Then, the entropy of $A$ has the highest value in this set of states only in the unique
stable equilibrium state $A_{\rm se1}=A_{\rm se}(E^A_1;\RA{})$
determined by $\RA{}$ and the value $E^A_1$. \dimostrazione{Theorem}{Theorem8}{Let $A_1$ be any state different from $A_{\rm se1}$  in the set of states of $A$ considered here. On account of Assumption \ref{Assumption2}  a zero work weight process
$\W{A}{1}{{\rm se1}}$ exists and is irreversible because a zero work weight process $\W{A}{{\rm se1}}{1}$
would violate the definition of stable equilibrium state.
Therefore,  Lemma \ref{Lemma7} implies $S^A_{\rm se1} >
S^A_1 $.}
\end{theorem}


\section{Fundamental relation, temperature, and
Gibbs relation for closed nonreactive systems}

In  this section we obtain, in our logical framework, some of the main standard theorems about properties of the stable equilibrium states of any closed nonreactive system.

\subsection*{Set of equivalent stable equilibrium states.} We
 call \textit{set of equivalent stable equilibrium states} of a
closed nonreactive system $A$, denoted $ESE^A$, a subset of its stable
equilibrium states such that any pair of states in the set: (1)
 differ from one another by some geometrical features of the regions of
space $\RA{}$; and (2) can be
interconnected by a zero-work reversible weight process for $A$
and, hence, by the definition of energy and Theorem \ref{Theorem6}, have the same  energy and the same
entropy.

\begin{comment} Let us recall that, for all the stable
equilibrium states of a closed system $A$ in a scenario $AB$,
system $A$ is separable and uncorrelated from its environment $B$. Moreover, for each of these states, the matter of $B$ does not produce force fields within $A$, so that the external force field $\FA$ is the stationary force field produced by sources external to the isolated system $AB$. Finally, since $A$ is closed and nonreactive, all the states of $A$ have the same composition.\end{comment}

\subsection*{Parameters of a closed system.} We  call
\textit{parameters} of a closed and nonreactive system $A$, denoted by
$\pmb{\beta}^A=\beta^A_1, \dots , \beta^A_s$, a minimal set of
real variables sufficient to fully and uniquely parametrize all
the different sets of equivalent stable equilibrium states $ESE^A$
of $A$. In the following, we  consider systems with a finite
number $s$ of parameters.

\spazio \begin{corollary}\label{Corollary10} \textbf{Fundamental relation for the stable equilibrium states of a closed system with no reactions}. On the set of all the
stable equilibrium states of a closed system $A$ (in scenario
$AB$) with a fixed composition $\textbf{\textit{n}}^{A}$, the entropy is related to the energy and the parameters by a single valued function
\begin{equation}\label{fundamental-relation}
S^A_{\rm se} = S^A_{\rm se}(E^A, \pmb{\beta}^A) \;\; ,
\end{equation}
which is called \textit{fundamental relation} for the stable
equilibrium states of $A$. \dimostrazione{Corollary}{Corollary10}{On account of Postulate \ref{Postulate2} and Lemma \ref{Lemma1},
among all the  states of a closed system $A$ with energy $E^A$,
the regions of space $\RA{}$ identify a unique stable equilibrium
state. This implies the existence of a single valued function
$A_{\rm se} = A_{\rm se}(E^A,\RA{})$, where $A_{\rm se}$ denotes
the state, in the sense of the vector containing the complete set of values of all the properties of the system. By definition, for
each value of the energy $E^A$, the values of the parameters
$\pmb{\beta}^A$ fully identify all the regions of space $\RA{}$
that correspond to a set of equivalent stable equilibrium states
$ESE^A$, which have the same value of the entropy. Therefore, the values of $E^A$ and $\pmb{\beta}^A$
fix uniquely the values of $S^A_{\rm se}$. This implies the existence of the
single valued functions written in Eq.
(\ref{fundamental-relation}).}
\end{corollary}


\begin{comment}  Usually \cite{HK,GB}, in view of the
equivalence that defines them, each set $ESE^A$ is thought of as a
single state called \virgolettedue{a stable equilibrium state} of $A$.
Thus, for a given closed system $A$ (and, hence,  given initial
amounts of constituents), it is commonly stated that the energy
and the parameters of $A$ determine \virgolettedue{a unique stable
equilibrium state} of $A$.\end{comment}

\begin{theorem}\label{Theorem10}  For any closed nonreactive system, for fixed
values of the parameters the fundamental relation
(\ref{fundamental-relation}) is a strictly increasing function of
the energy. \dimostrazione{Theorem}{Theorem10}{ Consider two stable equilibrium states
$A_{{\rm se}1}$ and $A_{{\rm se}2}$ of a closed system $A$, with
energies $E^A_1$ and $E^A_2$, entropies $S^A_{{\rm se}1}$ and
$S^A_{{\rm se}2}$, and with the same regions of space occupied by
the constituents of $A$ (and therefore the same values of the
parameters). Assume $E^A_2>E^A_1$. By Postulate \ref{Postulate3}, we can start
from state $A_{{\rm se}1}$ and, by a weight process for $A$ in
which the regions of space occupied by the constituents of $A$
have no net changes, add work so that the system ends in a
non-equilibrium state $A_2$ with energy $E^A_2$. By Theorem \ref{Theorem6}, we must have $S^A_2\ge S^A_{{\rm se}1}$. By Theorem \ref{Theorem8}, we have
$S^A_{{\rm se}2}> S^A_2$. Combining the two inequalities, we find
that $E^A_2>E^A_1$ implies $S^A_{{\rm se}2}> S^A_{{\rm se}1}$.}
\end{theorem}


 \begin{theorem}\label{Theorem11}
The fundamental relation (\ref{fundamental-relation}) is a differentiable function of
the energy. Moreover,
\begin{equation}\label{temperatureAsDerivative}\left.\frac{\partial S^A_{\rm se}(E^A, \pmb{\beta}^A)}{\partial E^A}\right|_{\pmb{\beta}^A} = \frac{1}{T^A(E^A, \RA{})} = \frac{1}{T^A(E^A, \pmb{\beta}^A)} \ ,
\end{equation}
where $\RA{}$ is any set of regions of space which corresponds to the set $\pmb{\beta}^A$ of the parameters. \dimostrazione{Theorem}{Theorem11}{Consider a stable equilibrium state $A_{{\rm se}1}$ of $A$, with regions of space $\RA{}$ and energy $E^A_{{\rm se}1}$, and a pair of states $(B_1, B_2)$ of any closed system $B$, such that in both states $B$ is separable and uncorrelated from its environment. Consider a reversible weight process for $BA$, from $B_1$ to $B_2$, standard with respect to $A$ and with initial state $A_{{\rm se}1}$ of $A$, and let $A_{{\rm se}2}$ be the final state of $A$ in this process. On account of the definition of entropy difference, Eq.\ (\ref{entropy}), one has
\begin{equation}\label{entropy-B}
S^B_2 - S^B_1 = - \int_{E^A_{\textrm{se}1}}^{E^A_{\textrm{se}2}}
\frac{1}{T^A(E^A;\RA{})} \; \textrm{d}E^A \;\; .
\end{equation}
Theorems \ref{Theorem6} and \ref{Theorem7} imply that, in the reversible weight process for $BA$ considered, the entropy change of $A$ is the opposite of that of $B$. Therefore, one has also
\begin{equation}\label{entropy-BA}
S^A_{{\rm se}2} - S^A_{{\rm se}1} = \int_{E^A_{\textrm{se}1}}^{E^A_{\textrm{se}2}}
\frac{1}{T^A(E^A;\RA{})} \; \textrm{d}E^A \;\; .
\end{equation}
Since, through proper choices of $(B_1, B_2)$, the final stable equilibrium state of $A$ can be changed arbitrarily, one has, for every stable equilibrium state of $A$ with the same regions of space $\RA{}$ as $A_{{\rm se}1}$
\begin{equation}\label{entropy-BAgeneral}
S^A_{{\rm se}} - S^A_{{\rm se}1} = \int_{E^A_{\textrm{se}1}}^{E^A_{\textrm{se}}}
\frac{1}{T^A(E^A;\RA{})} \; \textrm{d}E^A \;\; .
\end{equation}
On account of the fundamental theorem of integral calculus, the function $S^A_{\rm se} = S^A_{\rm se}(E^A, \RA{})$ defined by Eq.\ (\ref{entropy-BAgeneral}) is differentiable with respect to $E^A$ and its derivative is $1/T^A(E^A;\RA{})$. Since the function $S^A_{\rm se} = S^A_{\rm se}(E^A, \pmb{\beta}^A)$ defined by Eq.\ (\ref{fundamental-relation}) coincides with that defined by Eq.\ (\ref{entropy-BAgeneral}), it is differentiable with respect to $E^A$ as well, and its derivative is the same. Thus, Eq.\ (\ref{temperatureAsDerivative}) is proved.}
\end{theorem}

 \begin{corollary}\label{Corollary11} The fundamental relation $S^A_{\rm se} = S^A_{\rm se}(E^A, \pmb{\beta}^A)$ can be rewritten in the form
\begin{equation}\label{fundamental-inverse}
E^A_{\rm se} = E^A_{\rm se}(S^A, \pmb{\beta}^A) \;\; .
\end{equation}
The latter is differentiable with respect to $S_A$ and its derivative is given by
\begin{equation}\label{temperatureAsDerivative-inverse}\left.\frac{\partial E^A_{\rm se}(S^A, \pmb{\beta}^A)}{\partial S^A}\right|_{\pmb{\beta}^A} = T^A(E^A, \pmb{\beta}^A) \ .
\end{equation}
 \dimostrazione{Corollary}{Corollary11}{By Theorem \ref{Theorem10}, the fundamental relation $S^A_{\rm se} = S^A_{\rm se}(E^A, \pmb{\beta}^A)$ is a strictly increasing function of the energy. Therefore, it is invertible and yields the inverse function expressed by Eq.\ (\ref{fundamental-inverse}). Since the former is differentiable with respect to $E^A$, with derivative given by Eq.\ (\ref{temperatureAsDerivative}), the latter is differentiable with respect to $S^A$, and its derivative is given by Eq.\ (\ref{temperatureAsDerivative-inverse}).}
\end{corollary}


\subsection*{Fundamental relation in the quantum formalism.}
Let us recall that the measurement procedures that define energy
and entropy must be applied, in general, to an 
ensemble of identically prepared replicas of the system of
interest. Because the numerical outcomes may vary (fluctuate) from
replica to replica, the values of the energy and the entropy
defined by these procedures are arithmetic means. Therefore, what
we have denoted so far, for simplicity, by the symbols $E^A$ and
$S^A$ should be understood as $\langle E^A\rangle$ and $\langle
S^A\rangle$. Where appropriate, like in the quantum formalism
implementation, this more precise notation should be preferred.
Then, written in full notation, the fundamental relation
(\ref{fundamental-relation}) for a nonreactive closed system is
\begin{equation}\label{fundamental-relation-full}
\langle S^A\rangle_{\rm se} = S^A_{\rm se}(\langle E^A\rangle,
\pmb{\beta}^A) \;\; .
\end{equation}

 \begin{corollary}\label{Corollary12} \textbf{Gibbs relation for a non-reactive closed system}.
If the fundamental relation (\ref{fundamental-inverse}) is differentiable with
respect to each of the variables $\pmb{\beta}^A$, its  differential may be written as follows, where we omit the superscript \virgolettedue{$A$} and the subscript \virgolettedue{se} for simplicity,
\begin{equation}\label{Gibbs}
dE = T\, dS + \sum_{j\,=1}^s F_j \, d\beta_j \;\; ,
\end{equation}
where $F_j$ is called \textit{generalized force conjugated to the $j$-th
parameter of $A$}, $F_j= \big (\partial E_{\rm se}/\partial \beta_j
\big)_{S,\pmb{\beta}'}$. \dimostrazione{Corollary}{Corollary12}{The conclusion is a straightforward consequence of Corollary \ref{Corollary11} and of the assumption of differentiability with respect to $\pmb{\beta}^A$.}
\end{corollary}


\begin{comment} If all the regions of space $\RA{}$
coincide and the volume $V$ of any of them is a parameter, the
negative of the conjugated generalized force is called
\textit{pressure}, denoted by $p$, $p=- \big (\partial E_{\rm
se}/\partial V \big)_{S,\pmb{\beta}'}$. \end{comment}

\begin{theorem}\label{Theorem12} Consider two closed systems $A$ and
$B$, with fixed regions of space $\RA{}$ and $\RB{}$ occupied by
their constituents, with corresponding parameters $\pmb{\beta}^A$
and $\pmb{\beta}^B$. Then, the following are necessary conditions
for $A$ and $B$ to be in mutual stable equilibrium:
\begin{itemize}[noitemsep,nolistsep]
\item  their
states $A_1$ and $B_1$, with energy values $E^A_1$ and $E^B_1$
respectively, are stable equilibrium states;
\item the temperatures of $A$ and $B$ are equal, \textit{i.e.},
\begin{equation}\label{TemperatureEquality}T^A(E^A_1,\pmb{\beta}^A) =
T^B(E^B_1,\pmb{\beta}^B) \;\; ;
\end{equation}
\item there exists an interval of values of $\varepsilon$, centered in $\varepsilon=0$, such that
\begin{equation}\label{DifferentTemperatures1}
T^A(E^A_1+\varepsilon,\pmb{\beta}^A) <
T^B(E^B_1-\varepsilon,\pmb{\beta}^B) \;\; \textrm{if} \; \varepsilon < 0 \;\; ,
\end{equation}
\begin{equation}\label{DifferentTemperatures2}
T^A(E^A_1+\varepsilon,\pmb{\beta}^A) >
T^B(E^B_1-\varepsilon,\pmb{\beta}^B) \;\; \textrm{if} \; \varepsilon > 0 \;\; ,
\end{equation}
where, of course, $T(E,\pmb{\beta})$ denotes the inverse of $\partial S_{\rm
se}(E,\pmb{\beta})/\partial E$.
\end{itemize}
 \dimostrazione{Theorem}{Theorem12}{If either $A_1$ or $B_1$ were not a
stable equilibrium state, then by Assumption \ref{Assumption2} it could be changed to a different state in a zero-work weight process. Therefore, also
$C_1=A_1B_1$ could be changed to a different state with no
external effects; thus, it could not be a stable equilibrium
state.\\ Let us denote by $\Gamma^C(E^C_1)$ the set of all the
states of $C=AB$ such that: $A$ and $B$ are in stable equilibrium
states; the constituents of $A$ and $B$ are contained in the sets
of regions of space $\RA{}$ and $\RB{}$; the energy of $C$ has the
value $E^C_1=E^A_1+E^B_1$. On account of Theorem \ref{Theorem8}, a necessary
condition for $C_1$ to be a stable equilibrium state is that $C_1$
be the unique highest entropy state in the set $\Gamma^C(E^C_1)$.
By the additivity of entropy, we have
\begin{equation}\label{entropyadditivity-bis}
S^C = S^A + S^B \;\; .
\end{equation}
Because in the set $\Gamma^C(E^C_1)$ the states of $A$ and $B$ are
stable equilibrium, by Eq.\ (\ref{fundamental-relation}) we can
write $S^A=S_{\rm se}^A(E^A,\pmb{\beta}^A)$ and $S^B=S_{\rm
se}^B(E^B,\pmb{\beta}^B)$, where $\pmb{\beta}^A$ and
$\pmb{\beta}^B$ are the values of the parameters of $A$ and $B$
which correspond to the regions of space $\RA{}$ and $\RB{}$.
Moreover, since $E^A + E^B =E^C_1$, and $E^C_1=E^A_1+E^B_1$ is
fixed, $E^A=E^A_1+\varepsilon$, $E^B=E^B_1-\varepsilon$. Therefore, we may write
$S^C$ as
\begin{equation}\label{functionofEA}
S^C = S_{\rm se}^A(E^A_1+\varepsilon,\pmb{\beta}^A)+ S_{\rm
se}^B(E^B_1-\varepsilon,\pmb{\beta}^B)\;\; ,
\end{equation}
and, by differentiation with respect to $\varepsilon$, we readily
obtain
\begin{equation}\label{Derivative of SC}
\! \frac{\partial S^C}{\partial
\varepsilon}\!=\!
 \left( \frac{\partial S^A_{\rm
se}}{\partial E^A}\right)_{\pmbsub{\beta}^A}\!\!\! - \!  \left(
\frac{\partial S^B_{\rm se}}{\partial
E^B}\right)_{\pmbsub{\beta}^B} \!\!\!=\!
\frac{1}{T^A(E^A_1\!+\!\varepsilon,\pmb{\beta}^A)}\!-\!\frac{1}{T^B(E^B_1\!-\!\varepsilon,
\pmb{\beta}^B)}.
\end{equation}
Necessary conditions for $C_1$ (corresponding to $\varepsilon=0$)
to be the unique state which maximizes the entropy $S^C$ in the
set $\Gamma^C(E^C_1)$ are
\begin{equation}\label{HEcondition}
 \left.  \frac{\partial S^C}{\partial
 \varepsilon}\right|_{\varepsilon=0}  = 0 \;\;
 ,
\end{equation}
and,
\begin{equation}\label{HEcondition-two}
  \left.  \frac{\partial S^C}{\partial
 \varepsilon}\right|_{\varepsilon<0}  > 0 \;\; ,
 \;\;\;   \left.  \frac{\partial S^C}{\partial
 \varepsilon}\right|_{\varepsilon>0}  < 0 \;\; .
\end{equation}
Equations (\ref{Derivative of SC}) and (\ref{HEcondition}) prove
Eq.\ (\ref{TemperatureEquality}). Equations (\ref{Derivative of
SC}) and (\ref{HEcondition-two}) prove Eqs.
(\ref{DifferentTemperatures1}) and (\ref{DifferentTemperatures2}). }
\end{theorem}

\begin{postulate}\label{Postulate4} Any system $A$, in any stable
equilibrium state $A_{\rm{se}1}$, is in mutual stable equilibrium
with an identical copy $A^{\textrm{d}}$ of $A$, in the same state.\end{postulate}

\begin{comment} The statement of Postulate \ref{Postulate4} is usually considered as obvious in traditional treatments. However, it cannot be proved. Since the statement is useful to complete the treatment of the conditions for mutual stable equilibrium, it is here postulated explicitly. \end{comment}

\begin{corollary}\label{Corollary13} For the set of stable equilibrium
states of a system $A$ which correspond to a fixed set of values
of the parameters $\pmb{\beta}^A$, the temperature of $A$ is a
strictly increasing function of the energy of $A$. \dimostrazione{Corollary}{Corollary13}{Let $A_{\textrm{se}1}$ be any stable
equilibrium state of a closed system $A$, with regions of space
$\RA{}$ occupied by the constituents of $A$, which correspond to
the values $\pmb{\beta}^A$ of the parameters, and with an energy
value $E^A_1$. Let $A^{\textrm{d}}$ be an identical copy of $A$ and let
$A^{\textrm{d}}_{\textrm{se}1}$ be the stable equilibrium state of $A^{\textrm{d}}$
which is identical with $A_{\textrm{se}1}$. On account of
Postulate \ref{Postulate4}, $C_1 = A_{\textrm{se}1}A^{\textrm{d}}_{\textrm{se}1}$ is a
stable equilibrium state of $C = AA^{\textrm{d}}$. Therefore, by Theorem \ref{Theorem12}
and the fact that $A$ and  $A^{\textrm{d}}$, being identical, have identical
fundamental relations, there exists, in the neighborhood of $\varepsilon=0$, an interval of positive values of $\varepsilon$ such that
\begin{equation}\label{IncreasingTemperature}
T^A(E^A_1 \!+\! \varepsilon,\pmb{\beta}^A) > T^{A^{\textrm{d}}}(E^{A^{\textrm{d}}}_1 \!-\!
\varepsilon,\pmb{\beta}^A) = T^{A}(E^A_1 \!-\!
\varepsilon,\pmb{\beta}^A) \;\; .
\end{equation}
Therefore, in the neighborhood of $A_{\textrm{se}1}$, for fixed values
of the parameters $\pmb{\beta}^A$ the temperature of $A$ is a strictly increasing function of the energy of $A$. Since $A_{\textrm{se}1}$ has been chosen
arbitrarily, the conclusion is proved.}
\end{corollary}


\begin{comment} Corollary \ref{Corollary13} proves that the second
necessary condition for the mutual stable equilibrium of two
closed systems with fixed regions of space occupied by their
constituents, given by Eqs.\ (\ref{DifferentTemperatures1}) and (\ref{DifferentTemperatures2}), is
automatically fulfilled (once Postulate \ref{Postulate4} is stated, or implied).
For instance, for $\varepsilon>0$, using Eq.\ \ref{TemperatureEquality} and Corollary \ref{Corollary13}, we have
\begin{equation}\label{DifferentTemperatures2bis}
T^A(E^A_1+\varepsilon,\pmb{\beta}^A) > T^A_1=T^B_1>
T^B(E^B_1-\varepsilon,\pmb{\beta}^B) \;\; .
\end{equation}
\end{comment}

\begin{corollary}\label{Corollary14}
The existence of a system with two or more stable equilibrium states with the same regions of space occupied by the system, the same temperature, and different values of the energy is not allowed the basic laws of thermodynamics. \dimostrazione{Corollary}{Corollary14}{The conclusion is a direct consequence of Corollary \ref{Corollary13}.}
\end{corollary}


\begin{comment}
The usually adopted definition of a
thermal reservoir, which implies that it is a system with an infinite number of stable equilibrium states with the same regions of space occupied by its constituents, the same temperature, and different values of the energy, violates Corollary \ref{Corollary14} and is at best only an approximation of reality. Therefore, as already noted in Ref.\ \cite[p.87]{GB},  thermal reservoirs must be understood as \emph{idealized systems} whose defining features are \emph{limiting conditions that cannot exist on a strict basis, but are nevertheless approximated extremely well} by many practical macroscopic systems, such as a mole of a pure substance at the triple point or a system with a large enough mass that it can accommodate significant changes of energy with negligible changes of temperature. Instead, in the non-macroscopic realm (for example in the quantum thermodynamics framework \cite{Horodecki13}), we are not aware of few-particle model systems that exhibit the defining features of thermal reservoirs. \end{comment}

\begin{theorem}\label{Theorem13}  If two closed systems $A$ and
$B$ have fixed regions of space $\RA{}$ and $\RB{}$ occupied by
their constituents, then a necessary and sufficient condition for
them to be in mutual stable equilibrium is that their
states $A_1$ and $B_1$ be stable equilibrium states with the same
temperature, namely, $T^A_1 = T^B_1$. \dimostrazione{Theorem}{Theorem13}{By Theorem \ref{Theorem12}, the condition is necessary. We will now prove that it is sufficient. Let $A_1$ and $B_1$ be stable equilibrium states
of $A$ and $B$ with the same temperature, $T^A_1 = T^B_1$, and let
$E^C_1 = E^A_1 + E^B_1$ be  the energy of the composite system
$C$ in state $C_1 = A_1 B_1$.   Let us
assume, \textit{ab absurdo}, that in this state $A$ and $B$ are not in mutual stable equilibrium, i.e., $C_1$ is not a stable equilibrium state. On account of
Postulate \ref{Postulate2} and Lemma \ref{Lemma1}, for the given regions of space $\RA{}$ and $\RB{}$ and
the given energy value $E^C_1$, there exists a unique stable
equilibrium state of $C$. Since we assumed it is not $C_1$, let us denote it by  $C_2= A_2 B_2$, where necessarily $A_2$ and $B_2$ are stable equilibrium states. States $C_1$ and $C_2$ have the same energy, therefore, $E^A_2 + E^B_2=E^A_1 + E^B_1$. This means that if $E^A_2<E^A_1$, $E^B_2>E^B_1$. Then by Corollary 9 we have $T_2^A < T^A_1 = T^B_1 < T_2^B$ thus Eq.\ (\ref{TemperatureEquality}) of Theorem \ref{Theorem12} is not satisfied, $A$ and $B$ cannot be in mutual equilibrium, and so $C_2$ cannot be a stable equilibrium state. We reach a similar conclusion if $E^A_2>E^A_1$ because then $E^B_2<E^B_1$ and  Corollary 9 implies $T_2^A > T^A_1 = T^B_1 > T_2^B$. Therefore, we must have $E^A_2=E^A_1$ and $E^B_2=E^B_1$, but then, by Lemma 1, stable equilibrium state $A_2$ cannot be different from stable equilibrium state $A_1$, nor can $B_2$ be different from $B_1$. We are forced to conclude that, since $C_1$ coincides with $C_2$, it is a stable equilibrium state, contrary to the assumption.}
\end{theorem}

\section{Conclusions}

We presented a rigorous and general logical construction of an operational non-statistical definition of thermodynamic entropy which can be applied, \emph{in principle}, even to non-equilibrium states of few-particle systems, provided they are separable and uncorrelated from their environment. The new logical construction provides an operational definition of  entropy which requires neither the concept of \emph{heat} nor that of \emph{thermal reservoir}. Therefore, it removes: (1) the logical   limitations that restrict \emph{a priori} the traditional definitions of entropy to the equilibrium states of many-particle systems; (2) the operational limitations that restrict \emph{in practice} our previous definitions of non-equilibrium entropy  to many-particle systems; and  (3) the internal inconsistency of constructions that assume the existence of thermal reservoirs.

\end{document}